\documentclass[reprint,aps,prx,twocolumn,longbibliography,unsortedaddress]{revtex4-1}
\pdfoutput=1
\usepackage{graphicx}
\usepackage{epstopdf}
\usepackage{bm}
\usepackage{dcolumn}
\usepackage{amssymb}
\usepackage{amsmath}
\usepackage{gensymb}
\usepackage{array,multirow}
\usepackage{dcolumn}
\usepackage[normalem]{ulem}
\usepackage{cancel}
\usepackage{afterpage}
\usepackage{xfrac}
\usepackage{xcolor}
\usepackage[sort&compress]{natbib}
\usepackage[colorlinks=true,allcolors=blue]{hyperref}

\begin{document}

\title{Non-stoichiometry and Defects in the Weyl Semimetals TaAs, TaP, NbP, and NbAs}

\author{T. Besara,$^1$ D. Rhodes,$^1$ K.-W. Chen,$^1$ Q. Zhang,$^1$ B. Zeng,$^1$ Y. Xin,$^1$ L. Balicas,$^1$ R. E. Baumbach,$^1$ and T. Siegrist,$^{1,2}$}

\address{$^1$National High Magnetic Field Laboratory, Florida State University, Tallahassee, FL 32310, USA}
\address{$^2$Department of Chemical and Biomedical Engineering, FAMU-FSU College of Engineering, Florida State University, Tallahassee, FL 32310, USA}

\date{\today}

\begin{abstract}
We report a structural study of the Weyl semimetals TaAs, TaP, NbP, and NbAs, utilizing diffraction techniques (single crystal x-ray diffraction and energy dispersive spectroscopy) and imaging techniques (transmission electron microscopy/scanning transmission electron microscopy). We observe defects of various degrees, leading to non-stoichiometric single crystals of all four semimetals. While TaP displays a large pnictide deficiency with composition TaP$_{0.83(3)}$, and stacking faults accompanied by anti-site disorder and site vacancies, TaAs displays transition metal deficiency with composition Ta$_{0.92(2)}$As and a high density of stacking faults. NbP also displays pnictide deficiency, yielding composition NbP$_{0.95(2)}$, and lastly, NbAs display very little deviation from a 1:1 composition, NbAs$_{1.00(3)}$, and is therefore recommended to serve as the model compound for these semimetals.
\end{abstract}

\maketitle

\section{Introduction}
Weyl fermions, massless fermions predicted by Hermann Weyl in 1929~\cite{Weyl_ZP_1929} as solutions to the Dirac equation, have not yet been observed as fundamental particles in high energy physics. In 2011, however, it was predicted that Weyl fermions can be realized in condensed matter physics, as electronic quasi-particles in the family of pyrochlore iridates~\cite{Wan_PRB_2011} and the ferromagnetic spinel compound HgCr$_2$Se$_4$~\cite{Xu_PRL_2011}.

Following recent theoretical predictions of Weyl fermions in the simple semimetal TaAs and its isostructural compounds TaP, NbAs, and NbP~\cite{Weng_PRX_2015, Huang_NatComm_2015}, Weyl fermions were discovered experimentally shortly thereafter in TaAs~\cite{Xu_Science_2015, Lv_PRX_2015}. The discovery of Weyl fermions in TaAs was quickly confirmed by additional studies~\cite{Lv_NatPhys_2015, Yang_NatPhys_2015}, along with the discovery of Weyl fermions in the isostructural NbAs~\cite{Xu_NatPhys_2015} and TaP~\cite{Xu_SciAdv_2015}. Several studies on these semimetals have emerged: more detailed investigations of the Fermi surface topology~\cite{Lee_ARXIV_2015, Sun_PRB_2015, Lv_PRL_2015}, the observation of large magnetoresistance and high carrier mobility~\cite{Zhang_PRB_2015, Huang_PRX_2015, Shekhar_NatPhys_2015, Ghimire_JPCM_2015, Zhang_ARXIV_2015a, Shekhar_ARXIV_2015, Wang_ARXIV_2015}, the report of a quantum phase transition in TaP~\cite{Zhang_ARXIV_2015}, a Raman study of the lattice dynamics identifying all optical phonon modes in TaAs~\cite{Liu_PRB_2015}, a magnetization study of TaAs~\cite{Liu_ARXIV_2015}, and pressure studies of NbAs~\cite{Zhang_CPL_2015, Luo_ARXIV_2015}.\\
\\
Until now, all four semimetals, TaP, TaAs, NbP, and NbAs, have been studied with an assumed nominal 1:1 stoichiometric ratio between the transition metal and the pnictide. However, it is well known that the thermodynamic and transport properties of a material depend on the actual stoichiometry: e.g., the magnetoresistance and Fermi surface topology might be modified by disorder. In fact, a recent study observed quantum interference patterns arising from quasi-particle scattering near point defects on the surface of a single crystalline NbP~\cite{Zheng_ARXIV_2015}, followed by a theoretical investigation of surface state quasi-particle interference patterns in TaAs and NbP~\cite{Chang_ARXIV_2015}. It is therefore pertinent that a structural study be carried out on these materials to corroborate a 1:1 stoichiometric ratio that is independent of different sample batches, and to determine what, if any, defects are common.

A non-stoichiometric composition for one of these compounds was already reported in 1954~\cite{Schonberg_ACS_1954}: niobium phosphide was found as NbP$_{0.95}$, and an assumption was made that TaP would have a similar composition: TaP$_{0.95}$. In addition to the non-stoichiometry, it should be noted that the space group reported at the time was the centrosymmetric group $I4_{1}/amd$ for both materials. A decade later, the symmetry of the structure was corrected~\cite{Boller_ActaCryst_1963, Furuseth_ActaCryst_1964}, and it was shown that these isostructural semimetals in fact crystallize in the non-centrosymmetric space group $I4_{1}md$ (\#~109).

All four semimetals were extensively studied prior to the recent surge of interest. In addition to the already mentioned studies~\cite{Schonberg_ACS_1954, Boller_ActaCryst_1963, Furuseth_ActaCryst_1964}, the phase relations were explored in a number of reports~\cite{Furuseth_Nature_1964, Saini_CanJChem_1964, Furuseth_ActaChemScand_1964, Furuseth_ActaChemScand_1965, Rundqvist_Nature_1966, Murray_JLess_1976}. Willerstr\"{o}m~\cite{Willerstrom_JLess_1984}, in 1984, reported on stacking disorder in all four semimetals. The stacking disorder would originate from a formation of a metastable WC-type (hexagonal structure $P\bar{6}m2$) during the early stages of the synthesis reaction, that would partially transform into the stable NbAs-type structure, resulting in a structure of variable fractions of NbAs- and WC-type structures~\cite{Willerstrom_JLess_1984}. Depending on the temperature at which the powder samples were removed from the furnace, the nominal composition -- and the $c$-axis -- changed. Xu \emph{et al.}~\cite{Xu_InorgChem_1996}, in 1996, performed an extensive study on the crystal structure, electrical transport, and magnetic properties of single crystalline NbP. However, it was reported as stoichiometric with no deficiencies. In 2012 Saparov \emph{et al.}~\cite{Saparov_SST_2012} reported on extensive structure, thermodynamic and transport properties on a series of transition metal arsenides, including TaAs and NbAs. The samples, however, were not single crystals.\\
\\
Here, we report a structural study on this family of semimetals utilizing single crystal x-ray diffraction (XRD), energy dispersive spectroscopy (EDS), and transmission electron microscopy/scanning transmission electron microscopy (TEM/STEM). In addition, we present de Haas-van Alphen (magnetic torque) measurements on one of the members, TaP, to demonstrate that the specimens studied here are of similar quality to those discussed elsewhere.

XRD is essentially a measure of electron density and provides detailed information on the stoichiometry of a single crystal. However, it is not as sensitive to defects as other methods, as it assumes that all intensity is located in Bragg peaks. Therefore, point defects show up as reduced electron density, and stacking faults can result in apparent twinning or in the formation of anti-sites or anti-domains. Both of these will affect the electron density. EDS cannot give any information on possible defects, but it gives an approximate value of the composition of a single crystal and therefore the overall stoichiometry. However, it often has a large margin of error. In order to be able to detect defects, TEM and STEM can give a detailed picture of the atomic arrangement in a crystal, and, therefore, this technique can clearly identify any possible point and planar defects.

Utilizing these three methods, we show that these semimetals are, in fact, non-stoichiometric, and display noteworthy defect densities. The defects manifest themselves as site vacancies, anti-site disorder and anti-domain disorder due to stacking faults of layers of atoms. In TaP, we observe a pnictide deficiency accompanied by stacking faults, resulting in anti-site disorder and site vacancies. In contrast, for TaAs, we observe a transition metal deficiency in x-ray diffraction, accompanied by a high density of stacking faults only, no anti-site disorder or vacancies. NbP also displays pnictide deficiency, while in NbAs we observe very little or no deviation from the stoichiometric 1:1 ratio.

\begin{figure}[t]
    \begin{center}
        \includegraphics[width=1.0\columnwidth]{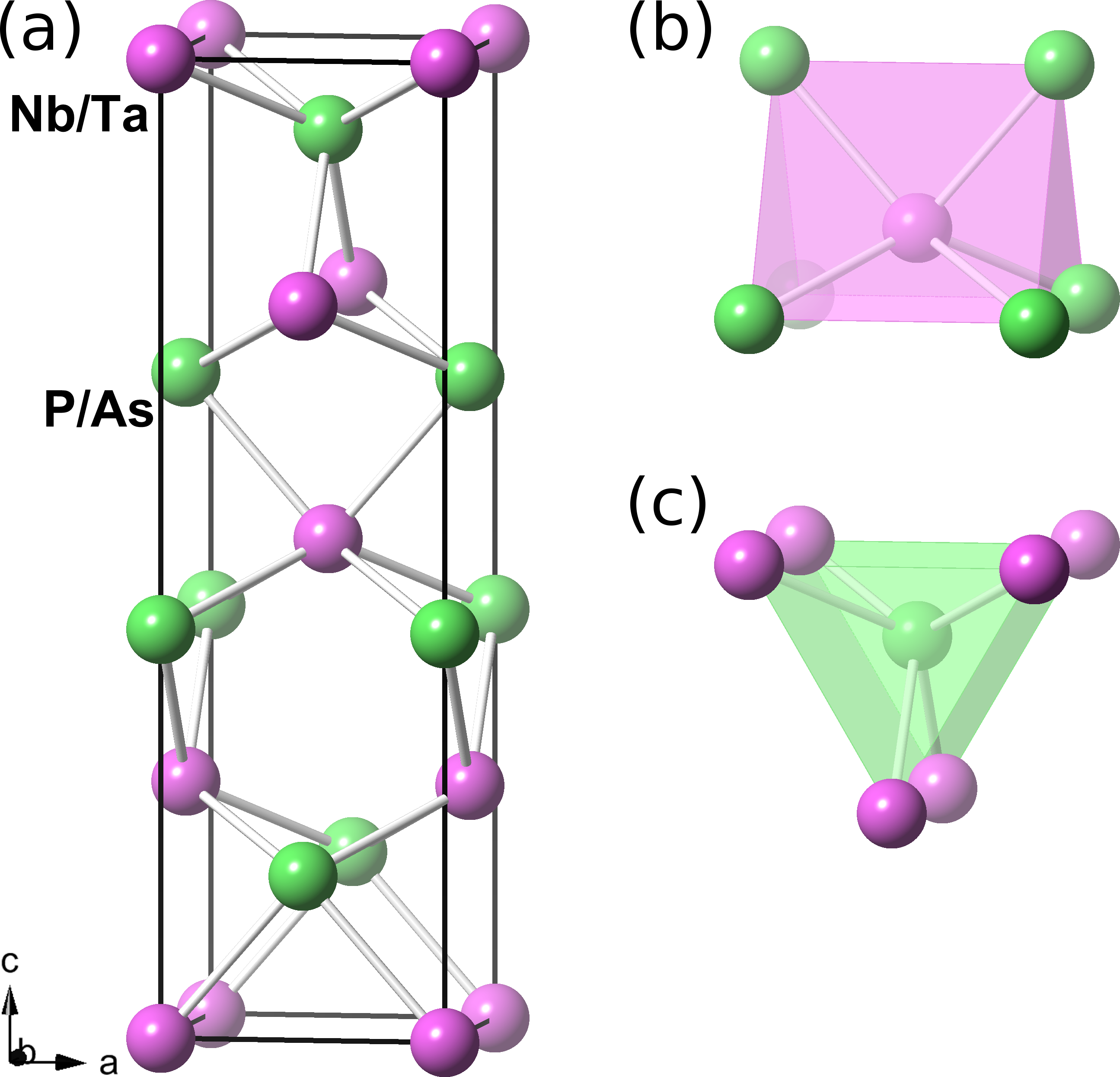}
        \caption{Crystal structure of \textit{TMPn} (\textit{TM}=Ta, Nb, and \textit{Pn}=As, P), which crystallizes in the non-centrosymmetric space group $I4_{1}md$, depicting (a) the unit cell, (b) the local coordination of the transition metal, and (c) the local coordination of the pnictide.}
        \label{fig:struct}
    \end{center}
\end{figure}

\section{Results \& Discussion}
The (Nb,Ta)(P,As) materials crystallize in the space group $I4_{1}md$ with the structure built up by a three dimensional network of trigonal prisms of \textit{TMPn}$_6$ and \textit{PnTM}$_6$ (\textit{TM}=transition metal Nb or Ta, \textit{Pn}=pnictide P or As), as can be seen in Fig.~\ref{fig:struct}. The results of the structural refinements from the single crystal x-ray diffraction of each semimetal are summarized in Table~\ref{tbl:xrd}. What follows are discussions of the results for each semimetal.\\
\\
\begin{table*}[t]
    \begin{center}
        \caption{Single crystal x-ray diffraction refinement parameters of the four semimetals, collected at ambient temperature. The semimetals crystallize in $I4_{1}md$ (\#109, $Z=4$). The bottom of the table lists the atomic parameters along with the anisotropic displacement parameters (in $\times10^{4}$ \textrm{\AA}$^2$). The atomic $x$- and $y$-coordinates are the same for all atoms. The transition metal (\emph{TM}) $z$-coordinate is at 4a ($0,0,z$) and is fixed at $z=0$, whereas the pnictide (\emph{Pn}) $z$-coordinate is refined.}
        \begin{tabular}{l c c c|c c|c c|c c|c c}
            \hline
            \hline
            \multicolumn{4}{l|}{} & \multicolumn{2}{c|}{\textbf{TaP}} & \multicolumn{2}{c|}{\textbf{TaAs}} & \multicolumn{2}{c|}{\textbf{NbP}} & \multicolumn{2}{c}{\textbf{NbAs}} \\
            \hline
            \hline
            \multicolumn{4}{l|}{\textbf{Composition}} & \multicolumn{2}{l|}{\textbf{TaP}$\mathbf{_{0.83(3)}}$} & \multicolumn{2}{l|}{\textbf{Ta}$\mathbf{_{0.92(2)}}$\textbf{As}} & \multicolumn{2}{l|}{\textbf{NbP}$\mathbf{_{0.95(2)}}$} & \multicolumn{2}{l}{\textbf{NbAs}$\mathbf{_{1.00(3)}}$} \\
            \multicolumn{4}{l|}{Formula weight (g/mol)} & \multicolumn{2}{l|}{206.76} & \multicolumn{2}{l|}{241.21} & \multicolumn{2}{l|}{122.47} & \multicolumn{2}{l}{167.80} \\
            \multicolumn{4}{l|}{$a$ (\textrm{\AA})} & \multicolumn{2}{l|}{3.31641(5)} & \multicolumn{2}{l|}{3.43646(7)} & \multicolumn{2}{l|}{3.33397(3)} & \multicolumn{2}{l}{3.45018(6)} \\
            \multicolumn{4}{l|}{$c$ (\textrm{\AA})} & \multicolumn{2}{l|}{11.3353(2)} & \multicolumn{2}{l|}{11.6417(3)} & \multicolumn{2}{l|}{11.3735(2)} & \multicolumn{2}{l}{11.6710(2)} \\
            \multicolumn{4}{l|}{$c/a$} & \multicolumn{2}{l|}{3.4179(1)} & \multicolumn{2}{l|}{3.3877(1)} & \multicolumn{2}{l|}{3.4114(1)} & \multicolumn{2}{l}{3.3827(1)} \\
            \multicolumn{4}{l|}{Volume (\textrm{\AA}$^{3}$)} & \multicolumn{2}{l|}{124.672(2)} & \multicolumn{2}{l|}{137.480(4)} & \multicolumn{2}{l|}{126.420(2)} & \multicolumn{2}{l}{138.928(3)} \\
            \multicolumn{4}{l|}{$\rho_{\textrm{calc}}$ (g/cm$^{3}$)} & \multicolumn{2}{l|}{11.015} & \multicolumn{2}{l|}{11.653} & \multicolumn{2}{l|}{6.434} & \multicolumn{2}{l}{8.022} \\
            \multicolumn{4}{l|}{Data collection range} & \multicolumn{2}{l|}{$6.41\degree\leq\theta\leq66.36\degree$} & \multicolumn{2}{l|}{$6.19\degree\leq\theta\leq66.22\degree$} & \multicolumn{2}{l|}{$6.38\degree\leq\theta\leq66.48\degree$} & \multicolumn{2}{l}{$6.17\degree\leq\theta\leq65.88\degree$} \\
            \multicolumn{4}{l|}{Reflections collected} & \multicolumn{2}{l|}{3409} & \multicolumn{2}{l|}{2562} & \multicolumn{2}{l|}{7919} & \multicolumn{2}{l}{3200} \\
            \multicolumn{4}{l|}{Independent reflections} & \multicolumn{2}{l|}{657} & \multicolumn{2}{l|}{714} & \multicolumn{2}{l|}{665} & \multicolumn{2}{l}{542} \\
            \multicolumn{4}{l|}{Parameters refined} & \multicolumn{2}{l|}{11} & \multicolumn{2}{l|}{11} & \multicolumn{2}{l|}{11} & \multicolumn{2}{l}{11} \\
            \multicolumn{4}{l|}{$R_{1}$, $wR_{2}$} & \multicolumn{2}{l|}{0.0543, 0.0971} & \multicolumn{2}{l|}{0.0527, 0.0984} & \multicolumn{2}{l|}{0.0477, 0.0798} & \multicolumn{2}{l}{0.0476, 0.0833} \\
            \multicolumn{4}{l|}{Goodness-of-fit on $F^{2}$} & \multicolumn{2}{l|}{0.9999} & \multicolumn{2}{l|}{1.0000} & \multicolumn{2}{l|}{0.9998} & \multicolumn{2}{l}{1.0000} \\
            \hline
            \multicolumn{4}{l|}{} & \multicolumn{2}{l|}{} & \multicolumn{2}{l|}{} & \multicolumn{2}{l|}{} & \multicolumn{2}{l}{} \\
            \hline
            Atom & Site & $x$ & $y$ & $z$ & $U_{\textrm{eq}}$ & $z$ & $U_{\textrm{eq}}$ & $z$ & $U_{\textrm{eq}}$ & $z$ & $U_{\textrm{eq}}$ \\
            \hline
            \emph{TM}=Ta, Nb & 4a & 0 & 0 & 0         & 30(4) & 0         & 15(3) & 0         & 19(2) & 0         & 26(6) \\
            \emph{Pn}=As, P  & 4a & 0 & 0 & 0.4173(3) & 16(2) & 0.4173(2) & 25(6) & 0.4176(2) & 24(6) & 0.4193(1) & 29(7) \\
            \hline
            \hline
        \end{tabular}
        \label{tbl:xrd}
    \end{center}
\end{table*}
\\
\textbf{TaP.} The first sample studied via XRD was the TaP. The single crystal refinement of the crystal structure displayed larger than expected anisotropic displacement parameters (ADPs) for phosphorus when compared to tantalum ADPs, indicating that the P-site could be deficient. Refining the site occupancy factor (SOF) of the P-site indeed yielded a significant drop in occupancy to 0.83, and a commensurate reduction of the anisotropic displacement parameters to match those of Ta. This scenario was repeated for a different crystal and yielded identical results within errors. Elemental analysis via EDS on a single crystal yielded a stoichiometry range of TaP$_{0.82-0.84}$, in excellent agreement with the XRD results.

\begin{figure*}[t]
    \begin{center}
        \includegraphics[width=1.0\textwidth]{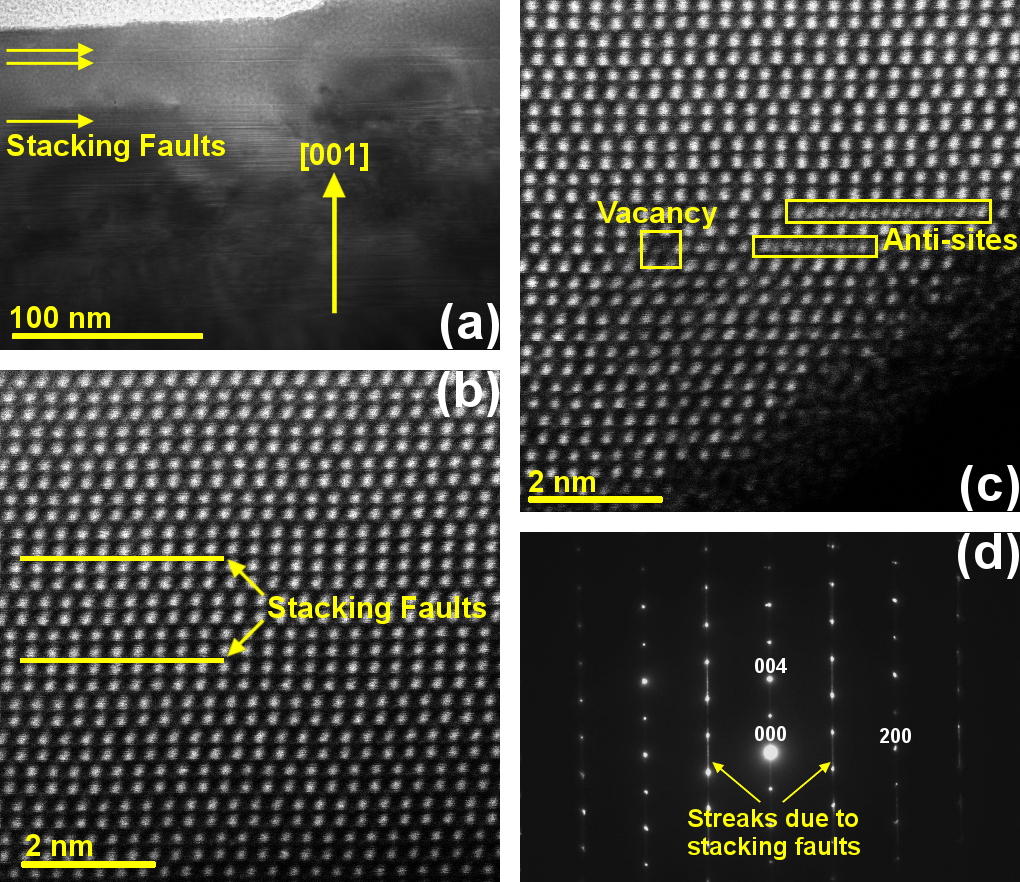}
        \caption{TEM/STEM images of TaP (the tilt in the images is due to drift). (a) TEM bright field image view of a TaP crystal with arrows highlighting the stacking faults. As can be seen, the faults are stacked along the $c$-axis. (b) Atomic resolution STEM HAADF Z-contrast image viewed along $[010]$, highlighting two stacking faults. (c) A STEM image highlighting anti-sites and a site vacancy. (d) Diffraction pattern of the $[010]$ direction showing streaks arising due to the stacking faults.}
        \label{fig:taptem}
    \end{center}
\end{figure*}

\begin{figure}[t]
    \begin{center}
        \includegraphics[width=1.0\columnwidth]{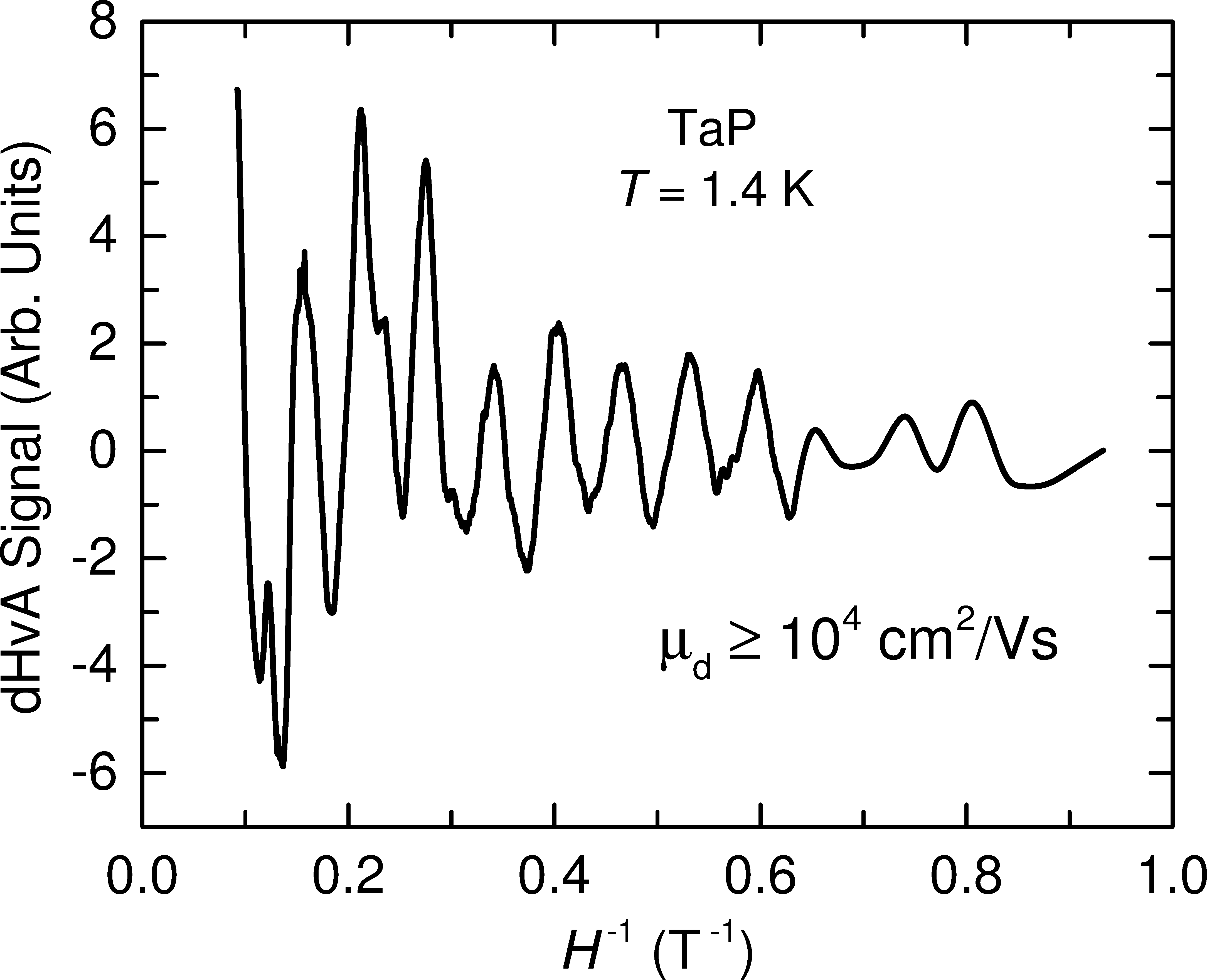}
        \caption{The de Haas-van Alphen signal (via magnetic torque measurements) of TaP. The signal has been normalized by the field and the background has been subtracted to reveal only the oscillatory signal.}
        \label{fig:tapdhva}
    \end{center}
\end{figure}

This non-stoichiometry hints at a large number of point defects (site vacancies) on the anionic P-sites, and could also be a sign of a potential superstructure, Ta$_6$P$_5$, in TaP, with 6 times the volume of the substructure. However, TEM/STEM images of a crushed crystal of TaP (Fig.~\ref{fig:taptem}) clearly reveal a somewhat different picture: defects in the form of stacking faults, anti-sites and site vacancies. In Fig.~\ref{fig:taptem}(a), the arrows indicate the stacking faults, always along the $c$-axis. Fig.~\ref{fig:taptem}(b) is a STEM high angle annular dark field (HAADF) $Z$-contrast image which shows the atomic arrangement. Qualitatively, the intensity of atomic columns in STEM HAADF images is proportional to the atomic number $Z^n$, where $n$ is close to 2, i.e., they are $Z$-number-sensitive images ($Z$-contrast). The bright spheres are the Ta-atoms which has $Z=73$. The atomic number of P is not high enough ($Z=15$) to show any appreciable intensity. The image clearly displays the alternating Ta- and P-layers, as expected from the structure, disrupted (the lines pointed to by arrows) by "shifts" of one half lattice width, creating regions with a different stacking arrangement. The TaAs STEM images, due to the larger atomic number of As, include a highlight of the arrangement (\emph{vide infra}). Fig.~\ref{fig:taptem}(c) is a STEM HAADF image of a different region, and it clearly displays anti-sites, i.e., areas where the P-sites are occupied by Ta and the Ta-sites are occupied by P. The Ta vacancy is also highlighted. Fig.~\ref{fig:taptem}(d) shows an electron diffraction pattern of the single crystal in the $(h0l)$ plane with streaks along the $c^*$ direction for $h\neq2n$. The streaks appear due to the stacking faults, and they are located in the $(h0l)$ plane with $h=2n+1$ positions (or, equivalently, in the $(0kl)$ with $k=2n+1$) due to the following reason: looking at the $ac$-plane of the crystal structure, one can identify "pairs" of Ta-atoms (or, equivalently, P-atoms) projected on the $ac$-plane and stacked along the $c$-axis. Disregarding the $b$-coordinate, one $(x,z)$ pair consists of the Ta in $(0,0)$ and the Ta in $(0,0.25)$. The next pair is stacked along the $c$-axis and is \emph{shifted} by half a lattice along the $a$-axis: one atom in $(0.5,0.5)$ and one in $(0.5,0.75)$. The following stacked pair is again shifted along the $a$-axis, and so on. When a stacking fault occurs, instead of a shift along the $a$-axis, the next pair is stacked directly above, or shifted along the $b$-axis (however, this is not visible in this $(h0l)$ plane diffraction pattern). To return to the original crystallization, an additional shift would be needed in order to bring the pairs into their "original" pattern. Therefore, since the stacking faults involve shifts by $\frac{1}{2}a$ (or equivalently, $\frac{1}{2}b$), only $(h0l)$ with $h$ odd show streaks.\\
\\
The clear evidence of defects in TaP may raise the question of crystal quality. However, magnetic torque measurements on a single crystal of TaP clearly display an oscillatory de Haas-van Alphen signal (Fig.~\ref{fig:tapdhva}). Oscillations are observed as low as 1 T, thus indicating a minimum drift mobility of $10^4$ cm$^2$/Vs at 1.4 K, comparable to other reports of ultrahigh carrier mobility in TaP~\cite{Shekhar_ARXIV_2015}. A Dingle plot (not shown here) suggests that the drift mobility exceeds $10^5$ cm$^2$/Vs, despite the large amount of defects. The observed quantum oscillations are a sign that, electronically, these crystals are of comparable quality to those reported elsewhere.\\
\\
\textbf{TaAs.} For TaAs, initially the same assumption was made as discovered in TaP, \emph{viz.} a pnictide deficiency. However, the single crystal refinement clearly showed that the As site is \emph{not} deficient. In fact, refining the As site occupancy factor (SOF), it increased above 1. Fixing the arsenic SOF at 1.0 while refining the tantalum SOF, the occupancy of the Ta site dropped to 0.92 (as expected), yielding a compound with the chemical formula Ta$_{0.92}$As. In contrast, EDS analysis on different single crystals yielded a wide range of elemental composition: from an As-deficient crystal (TaAs$_{0.91}$) to one with large Ta deficiency (Ta$_{0.7}$As). This wide range observed with EDS and the discrepancy between the XRD and EDS results can be understood by the existence of a high density of stacking faults, producing anti-domains. Different regions of crystals observed with EDS would show different stoichiometries depending on the density of stacking faults and defects.

\begin{figure*}[t]
    \begin{center}
        \includegraphics[width=1.0\textwidth]{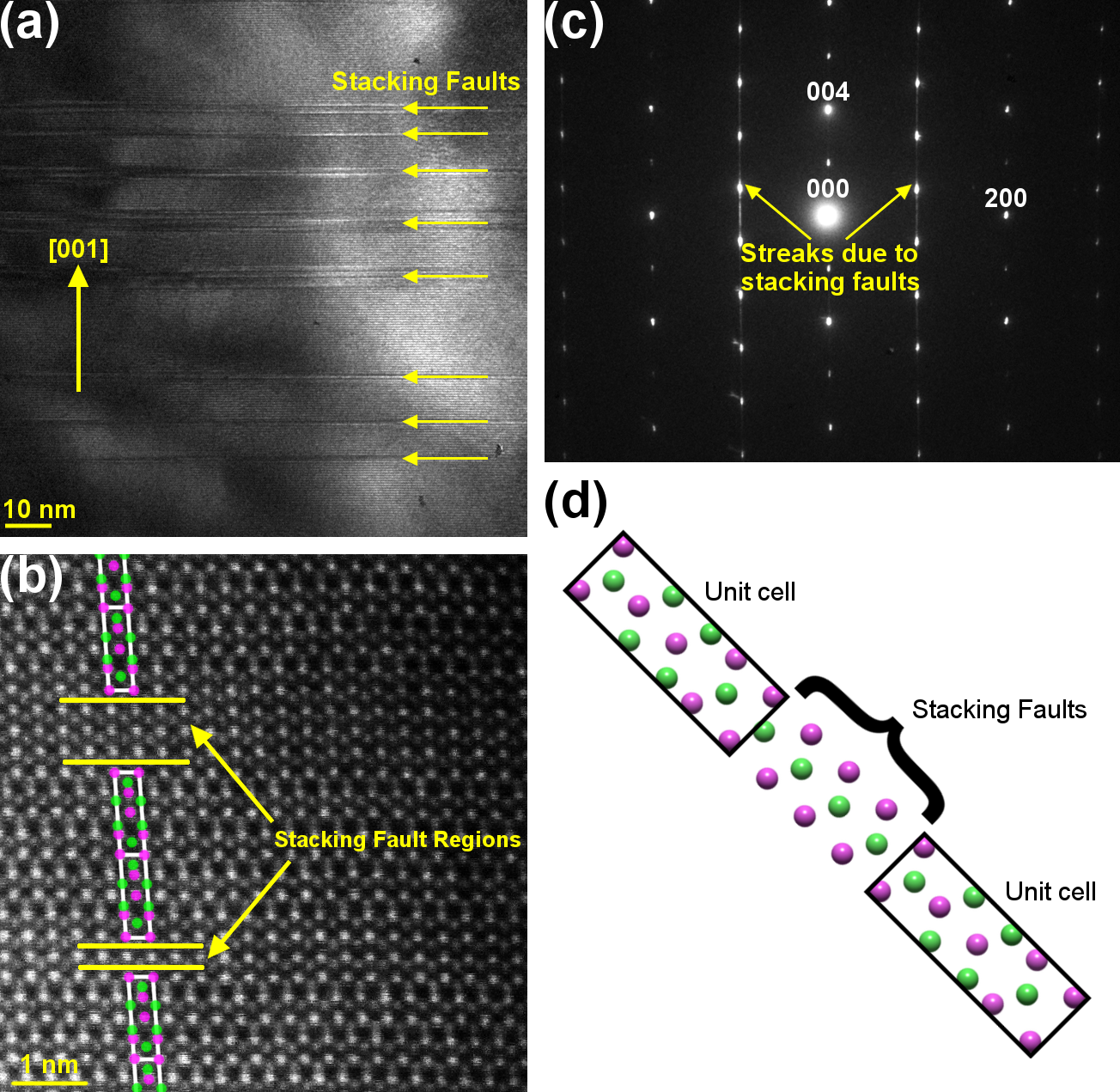}
        \caption{TEM/STEM images of TaAs (the tilt in the images is due to drift). (a) TEM bright field image view of a TaAs crystal. The faults in this case are also stacked along the $c$-axis. (b) Atomic resolution STEM HAADF Z-contrast image viewed along $[010]$. A strip has been highlighted to indicate unit cells of the known $I4_{1}md$ structure and two regions of stacking faults. (c) Diffraction pattern of the $[010]$ direction with streaks arising due to the stacking faults. (d) Illustration of two unit cells separated by a region of stacking faults.}
        \label{fig:taastem}
    \end{center}
\end{figure*}

TEM/STEM images of a crushed crystal of TaAs (Fig.~\ref{fig:taastem}) clearly show the high density of stacking faults. As in the case of TaP, the stacking faults are stacked along the $c$-axis (Fig.~\ref{fig:taastem}(a)). Fig.~\ref{fig:taastem}(b) show the STEM HAADF image where the Ta atoms are the bigger and brighter spheres. In this case, the atomic number of As ($Z=33$) is high enough to show an intensity and the As-atoms are the smaller and less bright ones. An area along the $c$-axis has been highlighted to show where the correct $I4_{1}md$ structure is broken up by \emph{regions} of stacking faults (between the lines as indicated by arrows). Fig.~\ref{fig:taastem}(c) shows the electron diffraction pattern of a TaAs single crystal, similarly to the TaP crystal, in the $(h0l)$ plane with streaks along the $c^*$ direction for $h\neq2n$. Fig.~\ref{fig:taastem}(d) is an illustration of the atomic arrangement which shows two unit cells of the correct $I4_{1}md$ structure separated by several layers of faulted stacking.\\
\\
\textbf{NbP.} The non-stoichiometric composition in this case, NbP$_{0.95(2)}$, was reported first in 1954~\cite{Schonberg_ACS_1954}, and our single crystal XRD refinement results corroborate that result. A second crystal yielded a different composition, NbP$_{0.928(13)}$, and this variation in composition was also confirmed by EDS measurements giving a compositional range from 0.93 to 0.95 P on one particular single crystal. This pnictide deficiency does not rule out stacking faults.\\
\\
\textbf{NbAs.} Unlike the other three semimetals in this family, NbAs did not display any major deviations from the 1:1 composition. Both XRD and EDS agree within estimated errors, showing nearly 1:1 Nb:As stoichiometry: XRD yielding NbAs$_{1.00(3)}$ while EDS showed a small range of $0.98-0.99$ As composition. Also here, the existence of stacking faults is not ruled out.

\section{Conclusion}
We have shown that the four semimetals TaAs, TaP, NbP, and NbAs clearly grow with an imperfect stoichiometry, and that at least two of these samples show large defects in the form of stacking faults, anti-site disordering, and/or site vacancies.

The differences between the four samples is striking: while TaP displays the full range of defects and grows with a large anionic pnictide deficiency, TaP$_{0.83(3)}$, TaAs shows only stacking faults, although at a higher density than TaP, accompanied by cationic transition metal deficiency, Ta$_{0.92(2)}$As. NbP is also pnictide-deficient, at a range comparable to what has already been reported~\cite{Schonberg_ACS_1954}, NbP$_{0.95(2)}$ which is closer to the "ideal" 1:1 composition. Lastly, NbAs has the nearly ideal 1:1 stoichiometric composition, although it observes small variations in the As occupancy, NbAs$_{1.00(3)}$.

Various experimental techniques are needed to characterize the defects. As an example, for TaAs, while XRD shows us Ta deficiencies in several crystals, EDS contrasts this by displaying a stoichiometry ranging from Ta deficient crystals to As deficient ones. TEM demonstrates the existence of a large number of stacking faults, stacked along the $c$-axis, and it became clear that the range in stoichiometry as given by EDS is due to the density of stacking faults in the scanned region of the single crystal.

Since NbAs deviated very little from a pure 1:1 stoichiometric compound, we suggest that NbAs may serve as the model compound for these Weyl semimetals.

Lastly, the magnetic torque measurements on TaP reveal oscillatory de Haas-van Alphen signals, and an ultrahigh carrier mobility. The existence of quantum oscillations has been previously taken as a proof of high quality crystals. Despite the high number of defects and large stoichiometry variations, it is clear that our crystals are electronically of comparable quality to those studied previously. The fact that quantum oscillations are observed in a single crystal with a high number of stacking faults and defects clearly illustrates the robustness of the quantum oscillations in these samples. Therefore, studies of these materials need to take into account the fact that structural defects that are clearly present and may shift the actual stoichiometry do not seem to couple to the electronic transport properties, reflected in the high mobilities observed. The observed one- and two-dimensional defects are expected to affect the band structure, and, hence, should be considered in calculations.

\section{Acknowledgments}
T.B. and T.S. are supported by the U.S. Department of Energy, Office of Basic Energy Sciences, Materials Sciences and Engineering Division, under Award \#DE-SC0008832. K.-W.C. and R.E.B. acknowledge support from the National High Magnetic Field Laboratory UCGP program. L.B. is supported by DOE-BES through Award \#DE-SC0002613. This work was performed at the National High Magnetic Field Laboratory, which is supported by the NSF cooperative agreement DMR-1157490 and the State of Florida.

\appendix
\section{Methods}
\textbf{Sample Preparation.} Single crystals of TaAs and NbAs were grown by chemical vapor transport, as previously described~\cite{Ghimire_JPCM_2015}. Polycrystalline precursor specimens were first prepared by sealing elemental Ta (Nb) and As mixtures under vacuum in quartz ampoules and heating the mixtures at a rate of 100 $\degree$C/hr to 700 $\degree$C, followed by a dwell at this temperature for 3 days. The polycrystalline TaAs and NbAs boules were subsequently sealed under vacuum in quartz ampoules with 3 mg/cm$^3$ of iodine to serve as the transporting agent. The ampoules had diameters 1.4 cm and lengths 10 cm and were placed in a horizontal tube furnace such that a temperature gradient would be established during firing. The ampoules were slowly heated at a rate of 18 $\degree$C/hr, reaching a temperature gradient of $\Delta T=850~\degree$C$-950~\degree$C. The ampoules were maintained under this condition for 3 weeks and were finally rapidly cooled to room temperature. This process produced a large number of single crystal specimens with typical dimensions of 0.5 mm on a side.

TaP and NP single crystals were synthesized through a chemical vapor transport technique using iodine as the transport agent. 99.99\% pure Nb (99.98\% pure Ta) powders, and 99\% pure P lumps were introduced into quartz tubes together with 99.999\% pure iodine serving as the transporting agent. The quartz tubes were vacuumed, brought to 500 $\degree$C, held at this temperature for 1 day, then brought up to 650 $\degree$C, held for half a day, and then finally raised to 975 $\degree$C and held there for 5 days. Subsequently, they were cooled to 800 $\degree$C, held there for 1 day, and were finally quenched in air.\\
\\
\textbf{Single Crystal X-ray Diffraction.} Crystals of the semimetals were structurally characterized by single crystal x-ray diffraction using an Oxford-Diffraction Xcalibur2 CCD system with graphite-monochromated Mo$K\alpha$ radiation. Data was collected to a resolution of 0.4~\textrm{\AA}, equivalent to $2\theta = 125\degree$. Reflections were recorded, indexed and corrected for absorption using the Agilent CrysAlisPro software~\cite{CrysAlisPro}. Subsequent structure refinements were carried out using CRYSTALS~\cite{Crystals}, using atomic positions from the literature~\cite{Pearson}. The data quality for all samples allowed for an unconstrained full matrix refinement against $F^2$, with anisotropic displacement parameters for all atoms. Crystallographic information files (CIFs) have been deposited with ICSD (CSD Nos. 430434, 430435, 430436, and 430437)~\cite{ICSD}.\\
\\
\textbf{EDS}. EDS was performed through field-emission scanning electron microscopy (Zeiss 1540 XB), on 6 to 12 spots of each of the several single crystals studied. The EDS stoichiometries quoted here results from average values.\\
\\
\textbf{TEM}. The TEM samples were prepared by crushing single crystals that were previously checked by XRD, in Ethyl Alcohol 200 Proof in a pestle and mortar. The suspension was then dropped onto a carbon/formvar TEM grid (Ted Pella, Inc.) using a 1.5 ml pipette. TEM/STEM images were collected using the probe aberration corrected JEOL JEM-ARM200cF with a cold field emission gun at 80 kV to avoid beam damage. The STEM high angle annular dark field (STEM-HAADF) images were taken with the JEOL HAADF detector using the following experimental conditions: probe size 7c, CL aperture 30 $\mu$m, scan speed 32 $\mu$s/pixel, and camera length 8 cm, which corresponds to a probe convergence semi-angle of 11 mrad and collection angles of $76-174.6$ mrad. The STEM resolution of the microscope is 0.78~\textrm{\AA}.

\bibliography{TMPn_bib}

%merlin.mbs apsrev4-1.bst 2010-07-25 4.21a (PWD, AO, DPC) hacked
%Control: key (0)
%Control: author (0) dotless jnrlst
%Control: editor formatted (1) identically to author
%Control: production of article title (0) allowed
%Control: page (1) range
%Control: year (0) verbatim
%Control: production of eprint (0) enabled
\begin{thebibliography}{44}%
\makeatletter
\providecommand \@ifxundefined [1]{%
 \@ifx{#1\undefined}
}%
\providecommand \@ifnum [1]{%
 \ifnum #1\expandafter \@firstoftwo
 \else \expandafter \@secondoftwo
 \fi
}%
\providecommand \@ifx [1]{%
 \ifx #1\expandafter \@firstoftwo
 \else \expandafter \@secondoftwo
 \fi
}%
\providecommand \natexlab [1]{#1}%
\providecommand \enquote  [1]{``#1''}%
\providecommand \bibnamefont  [1]{#1}%
\providecommand \bibfnamefont [1]{#1}%
\providecommand \citenamefont [1]{#1}%
\providecommand \href@noop [0]{\@secondoftwo}%
\providecommand \href [0]{\begingroup \@sanitize@url \@href}%
\providecommand \@href[1]{\@@startlink{#1}\@@href}%
\providecommand \@@href[1]{\endgroup#1\@@endlink}%
\providecommand \@sanitize@url [0]{\catcode `\\12\catcode `\$12\catcode
  `\&12\catcode `\#12\catcode `\^12\catcode `\_12\catcode `\%12\relax}%
\providecommand \@@startlink[1]{}%
\providecommand \@@endlink[0]{}%
\providecommand \url  [0]{\begingroup\@sanitize@url \@url }%
\providecommand \@url [1]{\endgroup\@href {#1}{\urlprefix }}%
\providecommand \urlprefix  [0]{URL }%
\providecommand \Eprint [0]{\href }%
\providecommand \doibase [0]{http://dx.doi.org/}%
\providecommand \selectlanguage [0]{\@gobble}%
\providecommand \bibinfo  [0]{\@secondoftwo}%
\providecommand \bibfield  [0]{\@secondoftwo}%
\providecommand \translation [1]{[#1]}%
\providecommand \BibitemOpen [0]{}%
\providecommand \bibitemStop [0]{}%
\providecommand \bibitemNoStop [0]{.\EOS\space}%
\providecommand \EOS [0]{\spacefactor3000\relax}%
\providecommand \BibitemShut  [1]{\csname bibitem#1\endcsname}%
\let\auto@bib@innerbib\@empty
%</preamble>
\bibitem [{\citenamefont {Weyl}(1929)}]{Weyl_ZP_1929}%
  \BibitemOpen
  \bibfield  {author} {\bibinfo {author} {\bibfnamefont {H.}~\bibnamefont
  {Weyl}},\ }\bibfield  {title} {\enquote {\bibinfo {title} {Elektron und
  {G}ravitation. {I}},}\ }\href {\doibase 10.1007/BF01339504} {\bibfield
  {journal} {\bibinfo  {journal} {Z. Phys.}\ }\textbf {\bibinfo {volume}
  {56}},\ \bibinfo {pages} {330--352} (\bibinfo {year} {1929})}\BibitemShut
  {NoStop}%
\bibitem [{\citenamefont {Wan}\ \emph {et~al.}(2011)\citenamefont {Wan},
  \citenamefont {Turner}, \citenamefont {Vishwanath},\ and\ \citenamefont
  {Savrasov}}]{Wan_PRB_2011}%
  \BibitemOpen
  \bibfield  {author} {\bibinfo {author} {\bibfnamefont {X.}~\bibnamefont
  {Wan}}, \bibinfo {author} {\bibfnamefont {A.~M.}\ \bibnamefont {Turner}},
  \bibinfo {author} {\bibfnamefont {A.}~\bibnamefont {Vishwanath}}, \ and\
  \bibinfo {author} {\bibfnamefont {S.~Y.}\ \bibnamefont {Savrasov}},\
  }\bibfield  {title} {\enquote {\bibinfo {title} {Topological semimetal and
  {F}ermi-arc surface states in the electronic structure of pyrochlore
  iridates},}\ }\href {\doibase 10.1103/PhysRevB.83.205101} {\bibfield
  {journal} {\bibinfo  {journal} {Phys. Rev. B}\ }\textbf {\bibinfo {volume}
  {83}},\ \bibinfo {pages} {205101} (\bibinfo {year} {2011})}\BibitemShut
  {NoStop}%
\bibitem [{\citenamefont {Xu}\ \emph {et~al.}(2011)\citenamefont {Xu},
  \citenamefont {Weng}, \citenamefont {Wang}, \citenamefont {Dai},\ and\
  \citenamefont {Fang}}]{Xu_PRL_2011}%
  \BibitemOpen
  \bibfield  {author} {\bibinfo {author} {\bibfnamefont {G.}~\bibnamefont
  {Xu}}, \bibinfo {author} {\bibfnamefont {H.}~\bibnamefont {Weng}}, \bibinfo
  {author} {\bibfnamefont {Z.}~\bibnamefont {Wang}}, \bibinfo {author}
  {\bibfnamefont {X.}~\bibnamefont {Dai}}, \ and\ \bibinfo {author}
  {\bibfnamefont {Z.}~\bibnamefont {Fang}},\ }\bibfield  {title} {\enquote
  {\bibinfo {title} {Chern semimetal and the quantized anomalous {H}all effect
  in {H}g{C}r$_2${S}e$_4$},}\ }\href {\doibase 10.1103/PhysRevLett.107.186806}
  {\bibfield  {journal} {\bibinfo  {journal} {Phys. Rev. Lett.}\ }\textbf
  {\bibinfo {volume} {107}},\ \bibinfo {pages} {186806} (\bibinfo {year}
  {2011})}\BibitemShut {NoStop}%
\bibitem [{\citenamefont {Weng}\ \emph {et~al.}(2015)\citenamefont {Weng},
  \citenamefont {Fang}, \citenamefont {Fang}, \citenamefont {Bernevig},\ and\
  \citenamefont {Dai}}]{Weng_PRX_2015}%
  \BibitemOpen
  \bibfield  {author} {\bibinfo {author} {\bibfnamefont {H.}~\bibnamefont
  {Weng}}, \bibinfo {author} {\bibfnamefont {C.}~\bibnamefont {Fang}}, \bibinfo
  {author} {\bibfnamefont {Z.}~\bibnamefont {Fang}}, \bibinfo {author}
  {\bibfnamefont {B.~A.}\ \bibnamefont {Bernevig}}, \ and\ \bibinfo {author}
  {\bibfnamefont {X.}~\bibnamefont {Dai}},\ }\bibfield  {title} {\enquote
  {\bibinfo {title} {{W}eyl semimetal phase in noncentrosymmetric
  transition-metal monophosphides},}\ }\href {\doibase
  10.1103/PhysRevX.5.011029} {\bibfield  {journal} {\bibinfo  {journal} {Phys.
  Rev. X}\ }\textbf {\bibinfo {volume} {5}},\ \bibinfo {pages} {011029}
  (\bibinfo {year} {2015})}\BibitemShut {NoStop}%
\bibitem [{\citenamefont {Huang}\ \emph
  {et~al.}(2015{\natexlab{a}})\citenamefont {Huang}, \citenamefont {Xu},
  \citenamefont {Belopolski}, \citenamefont {Lee}, \citenamefont {Chang},
  \citenamefont {Wang}, \citenamefont {Alidoust}, \citenamefont {Bian},
  \citenamefont {Neupane}, \citenamefont {Zhang}, \citenamefont {Jia},
  \citenamefont {Bansil}, \citenamefont {Lin},\ and\ \citenamefont
  {Hasan}}]{Huang_NatComm_2015}%
  \BibitemOpen
  \bibfield  {author} {\bibinfo {author} {\bibfnamefont {S.-M.}\ \bibnamefont
  {Huang}}, \bibinfo {author} {\bibfnamefont {S.-Y.}\ \bibnamefont {Xu}},
  \bibinfo {author} {\bibfnamefont {I.}~\bibnamefont {Belopolski}}, \bibinfo
  {author} {\bibfnamefont {C.-C.}\ \bibnamefont {Lee}}, \bibinfo {author}
  {\bibfnamefont {G.}~\bibnamefont {Chang}}, \bibinfo {author} {\bibfnamefont
  {B.}~\bibnamefont {Wang}}, \bibinfo {author} {\bibfnamefont {N.}~\bibnamefont
  {Alidoust}}, \bibinfo {author} {\bibfnamefont {G.}~\bibnamefont {Bian}},
  \bibinfo {author} {\bibfnamefont {M.}~\bibnamefont {Neupane}}, \bibinfo
  {author} {\bibfnamefont {C.}~\bibnamefont {Zhang}}, \bibinfo {author}
  {\bibfnamefont {S.}~\bibnamefont {Jia}}, \bibinfo {author} {\bibfnamefont
  {A.}~\bibnamefont {Bansil}}, \bibinfo {author} {\bibfnamefont
  {H.}~\bibnamefont {Lin}}, \ and\ \bibinfo {author} {\bibfnamefont {M.~Z.}\
  \bibnamefont {Hasan}},\ }\bibfield  {title} {\enquote {\bibinfo {title} {A
  {W}eyl {F}ermion semimetal with surface {F}ermi arcs in the transition metal
  monopnictide {T}a{A}s class},}\ }\href {\doibase 10.1038/ncomms8373}
  {\bibfield  {journal} {\bibinfo  {journal} {Nat. Commun.}\ }\textbf {\bibinfo
  {volume} {6}},\ \bibinfo {pages} {7373} (\bibinfo {year}
  {2015}{\natexlab{a}})}\BibitemShut {NoStop}%
\bibitem [{\citenamefont {Xu}\ \emph {et~al.}(2015{\natexlab{a}})\citenamefont
  {Xu}, \citenamefont {Belopolski}, \citenamefont {Alidoust}, \citenamefont
  {Neupane}, \citenamefont {Bian}, \citenamefont {Zhang}, \citenamefont
  {Sankar}, \citenamefont {Chang}, \citenamefont {Yuan}, \citenamefont {Lee},
  \citenamefont {Huang}, \citenamefont {Zheng}, \citenamefont {Ma},
  \citenamefont {Sanchez}, \citenamefont {Wang}, \citenamefont {Bansil},
  \citenamefont {Chou}, \citenamefont {Shibayev}, \citenamefont {Lin},
  \citenamefont {Jia},\ and\ \citenamefont {Hasan}}]{Xu_Science_2015}%
  \BibitemOpen
  \bibfield  {author} {\bibinfo {author} {\bibfnamefont {S.-Y.}\ \bibnamefont
  {Xu}}, \bibinfo {author} {\bibfnamefont {I.}~\bibnamefont {Belopolski}},
  \bibinfo {author} {\bibfnamefont {N.}~\bibnamefont {Alidoust}}, \bibinfo
  {author} {\bibfnamefont {M.}~\bibnamefont {Neupane}}, \bibinfo {author}
  {\bibfnamefont {G.}~\bibnamefont {Bian}}, \bibinfo {author} {\bibfnamefont
  {C.}~\bibnamefont {Zhang}}, \bibinfo {author} {\bibfnamefont
  {R.}~\bibnamefont {Sankar}}, \bibinfo {author} {\bibfnamefont
  {G.}~\bibnamefont {Chang}}, \bibinfo {author} {\bibfnamefont
  {Z.}~\bibnamefont {Yuan}}, \bibinfo {author} {\bibfnamefont {C.-C.}\
  \bibnamefont {Lee}}, \bibinfo {author} {\bibfnamefont {S.-M.}\ \bibnamefont
  {Huang}}, \bibinfo {author} {\bibfnamefont {H.}~\bibnamefont {Zheng}},
  \bibinfo {author} {\bibfnamefont {J.}~\bibnamefont {Ma}}, \bibinfo {author}
  {\bibfnamefont {D.~S.}\ \bibnamefont {Sanchez}}, \bibinfo {author}
  {\bibfnamefont {B.}~\bibnamefont {Wang}}, \bibinfo {author} {\bibfnamefont
  {A.}~\bibnamefont {Bansil}}, \bibinfo {author} {\bibfnamefont
  {F.}~\bibnamefont {Chou}}, \bibinfo {author} {\bibfnamefont {P.~P.}\
  \bibnamefont {Shibayev}}, \bibinfo {author} {\bibfnamefont {H.}~\bibnamefont
  {Lin}}, \bibinfo {author} {\bibfnamefont {S.}~\bibnamefont {Jia}}, \ and\
  \bibinfo {author} {\bibfnamefont {M.~Z.}\ \bibnamefont {Hasan}},\ }\bibfield
  {title} {\enquote {\bibinfo {title} {Discovery of a {W}eyl fermion semimetal
  and topological {F}ermi arcs},}\ }\href {\doibase 10.1126/science.aaa9297}
  {\bibfield  {journal} {\bibinfo  {journal} {Science}\ }\textbf {\bibinfo
  {volume} {349}},\ \bibinfo {pages} {613--617} (\bibinfo {year}
  {2015}{\natexlab{a}})}\BibitemShut {NoStop}%
\bibitem [{\citenamefont {Lv}\ \emph {et~al.}(2015{\natexlab{a}})\citenamefont
  {Lv}, \citenamefont {Weng}, \citenamefont {Fu}, \citenamefont {Wang},
  \citenamefont {Miao}, \citenamefont {Ma}, \citenamefont {Richard},
  \citenamefont {Huang}, \citenamefont {Zhao}, \citenamefont {Chen},
  \citenamefont {Fang}, \citenamefont {Dai}, \citenamefont {Qian},\ and\
  \citenamefont {Ding}}]{Lv_PRX_2015}%
  \BibitemOpen
  \bibfield  {author} {\bibinfo {author} {\bibfnamefont {B.~Q.}\ \bibnamefont
  {Lv}}, \bibinfo {author} {\bibfnamefont {H.~M.}\ \bibnamefont {Weng}},
  \bibinfo {author} {\bibfnamefont {B.~B.}\ \bibnamefont {Fu}}, \bibinfo
  {author} {\bibfnamefont {X.~P.}\ \bibnamefont {Wang}}, \bibinfo {author}
  {\bibfnamefont {H.}~\bibnamefont {Miao}}, \bibinfo {author} {\bibfnamefont
  {J.}~\bibnamefont {Ma}}, \bibinfo {author} {\bibfnamefont {P.}~\bibnamefont
  {Richard}}, \bibinfo {author} {\bibfnamefont {X.~C.}\ \bibnamefont {Huang}},
  \bibinfo {author} {\bibfnamefont {L.~X.}\ \bibnamefont {Zhao}}, \bibinfo
  {author} {\bibfnamefont {G.~F.}\ \bibnamefont {Chen}}, \bibinfo {author}
  {\bibfnamefont {Z.}~\bibnamefont {Fang}}, \bibinfo {author} {\bibfnamefont
  {X.}~\bibnamefont {Dai}}, \bibinfo {author} {\bibfnamefont {T.}~\bibnamefont
  {Qian}}, \ and\ \bibinfo {author} {\bibfnamefont {H.}~\bibnamefont {Ding}},\
  }\bibfield  {title} {\enquote {\bibinfo {title} {Experimental discovery of
  {W}eyl semimetal {T}a{A}s},}\ }\href {\doibase 10.1103/PhysRevX.5.031013}
  {\bibfield  {journal} {\bibinfo  {journal} {Phys. Rev. X}\ }\textbf {\bibinfo
  {volume} {5}},\ \bibinfo {pages} {031013} (\bibinfo {year}
  {2015}{\natexlab{a}})}\BibitemShut {NoStop}%
\bibitem [{\citenamefont {Lv}\ \emph {et~al.}(2015{\natexlab{b}})\citenamefont
  {Lv}, \citenamefont {Xu}, \citenamefont {Weng}, \citenamefont {Ma},
  \citenamefont {Richard}, \citenamefont {Huang}, \citenamefont {Zhao},
  \citenamefont {Chen}, \citenamefont {Matt}, \citenamefont {Bisti},
  \citenamefont {Strocov}, \citenamefont {Mesot}, \citenamefont {Fang},
  \citenamefont {Dai}, \citenamefont {Qian}, \citenamefont {Shi},\ and\
  \citenamefont {Ding}}]{Lv_NatPhys_2015}%
  \BibitemOpen
  \bibfield  {author} {\bibinfo {author} {\bibfnamefont {B.~Q.}\ \bibnamefont
  {Lv}}, \bibinfo {author} {\bibfnamefont {N.}~\bibnamefont {Xu}}, \bibinfo
  {author} {\bibfnamefont {H.~M.}\ \bibnamefont {Weng}}, \bibinfo {author}
  {\bibfnamefont {J.~Z.}\ \bibnamefont {Ma}}, \bibinfo {author} {\bibfnamefont
  {P.}~\bibnamefont {Richard}}, \bibinfo {author} {\bibfnamefont {X.~C.}\
  \bibnamefont {Huang}}, \bibinfo {author} {\bibfnamefont {L.~X.}\ \bibnamefont
  {Zhao}}, \bibinfo {author} {\bibfnamefont {G.~F.}\ \bibnamefont {Chen}},
  \bibinfo {author} {\bibfnamefont {C.~E.}\ \bibnamefont {Matt}}, \bibinfo
  {author} {\bibfnamefont {F.}~\bibnamefont {Bisti}}, \bibinfo {author}
  {\bibfnamefont {V.~N.}\ \bibnamefont {Strocov}}, \bibinfo {author}
  {\bibfnamefont {J.}~\bibnamefont {Mesot}}, \bibinfo {author} {\bibfnamefont
  {Z.}~\bibnamefont {Fang}}, \bibinfo {author} {\bibfnamefont {X.}~\bibnamefont
  {Dai}}, \bibinfo {author} {\bibfnamefont {T.}~\bibnamefont {Qian}}, \bibinfo
  {author} {\bibfnamefont {M.}~\bibnamefont {Shi}}, \ and\ \bibinfo {author}
  {\bibfnamefont {H.}~\bibnamefont {Ding}},\ }\bibfield  {title} {\enquote
  {\bibinfo {title} {Observation of {W}eyl nodes in {T}a{A}s},}\ }\href
  {\doibase 10.1038/nphys3426} {\bibfield  {journal} {\bibinfo  {journal} {Nat.
  Phys.}\ }\textbf {\bibinfo {volume} {11}} (\bibinfo {year}
  {2015}{\natexlab{b}}),\ 10.1038/nphys3426}\BibitemShut {NoStop}%
\bibitem [{\citenamefont {Yang}\ \emph {et~al.}(2015)\citenamefont {Yang},
  \citenamefont {Liu}, \citenamefont {Sun}, \citenamefont {Peng}, \citenamefont
  {Yang}, \citenamefont {Zhang}, \citenamefont {Zhou}, \citenamefont {Zhang},
  \citenamefont {Guo}, \citenamefont {Rahn}, \citenamefont {Prabhakaran},
  \citenamefont {Hussain}, \citenamefont {Mo}, \citenamefont {Felser},
  \citenamefont {Yan},\ and\ \citenamefont {Chen}}]{Yang_NatPhys_2015}%
  \BibitemOpen
  \bibfield  {author} {\bibinfo {author} {\bibfnamefont {L.~X.}\ \bibnamefont
  {Yang}}, \bibinfo {author} {\bibfnamefont {Z.~K.}\ \bibnamefont {Liu}},
  \bibinfo {author} {\bibfnamefont {Y.}~\bibnamefont {Sun}}, \bibinfo {author}
  {\bibfnamefont {H.}~\bibnamefont {Peng}}, \bibinfo {author} {\bibfnamefont
  {H.~F.}\ \bibnamefont {Yang}}, \bibinfo {author} {\bibfnamefont
  {T.}~\bibnamefont {Zhang}}, \bibinfo {author} {\bibfnamefont
  {B.}~\bibnamefont {Zhou}}, \bibinfo {author} {\bibfnamefont {Y.}~\bibnamefont
  {Zhang}}, \bibinfo {author} {\bibfnamefont {Y.~F.}\ \bibnamefont {Guo}},
  \bibinfo {author} {\bibfnamefont {M.}~\bibnamefont {Rahn}}, \bibinfo {author}
  {\bibfnamefont {D.}~\bibnamefont {Prabhakaran}}, \bibinfo {author}
  {\bibfnamefont {Z.}~\bibnamefont {Hussain}}, \bibinfo {author} {\bibfnamefont
  {S.~K.}\ \bibnamefont {Mo}}, \bibinfo {author} {\bibfnamefont
  {C.}~\bibnamefont {Felser}}, \bibinfo {author} {\bibfnamefont
  {B.}~\bibnamefont {Yan}}, \ and\ \bibinfo {author} {\bibfnamefont {Y.~L.}\
  \bibnamefont {Chen}},\ }\bibfield  {title} {\enquote {\bibinfo {title}
  {{W}eyl semimetal phase in the non-centrosymmetric compound {T}a{A}s},}\
  }\href {\doibase 10.1038/nphys3425} {\bibfield  {journal} {\bibinfo
  {journal} {Nat. Phys.}\ }\textbf {\bibinfo {volume} {11}},\ \bibinfo {pages}
  {728--732} (\bibinfo {year} {2015})}\BibitemShut {NoStop}%
\bibitem [{\citenamefont {Xu}\ \emph {et~al.}(2015{\natexlab{b}})\citenamefont
  {Xu}, \citenamefont {Alidoust}, \citenamefont {Belopolski}, \citenamefont
  {Yuan}, \citenamefont {Bian}, \citenamefont {Chang}, \citenamefont {Zheng},
  \citenamefont {Strocov}, \citenamefont {Sanchez}, \citenamefont {Chang},
  \citenamefont {Zhang}, \citenamefont {Mou}, \citenamefont {Wu}, \citenamefont
  {Huang}, \citenamefont {Lee}, \citenamefont {Huang}, \citenamefont {Wang},
  \citenamefont {Bansil}, \citenamefont {Jeng}, \citenamefont {Neupert},
  \citenamefont {Kaminski}, \citenamefont {Lin}, \citenamefont {Jia},\ and\
  \citenamefont {Hasan}}]{Xu_NatPhys_2015}%
  \BibitemOpen
  \bibfield  {author} {\bibinfo {author} {\bibfnamefont {S.-Y.}\ \bibnamefont
  {Xu}}, \bibinfo {author} {\bibfnamefont {N.}~\bibnamefont {Alidoust}},
  \bibinfo {author} {\bibfnamefont {I.}~\bibnamefont {Belopolski}}, \bibinfo
  {author} {\bibfnamefont {Z.}~\bibnamefont {Yuan}}, \bibinfo {author}
  {\bibfnamefont {G.}~\bibnamefont {Bian}}, \bibinfo {author} {\bibfnamefont
  {T.-R.}\ \bibnamefont {Chang}}, \bibinfo {author} {\bibfnamefont
  {H.}~\bibnamefont {Zheng}}, \bibinfo {author} {\bibfnamefont {V.~N.}\
  \bibnamefont {Strocov}}, \bibinfo {author} {\bibfnamefont {D.~S.}\
  \bibnamefont {Sanchez}}, \bibinfo {author} {\bibfnamefont {G.}~\bibnamefont
  {Chang}}, \bibinfo {author} {\bibfnamefont {C.}~\bibnamefont {Zhang}},
  \bibinfo {author} {\bibfnamefont {D.}~\bibnamefont {Mou}}, \bibinfo {author}
  {\bibfnamefont {Y.}~\bibnamefont {Wu}}, \bibinfo {author} {\bibfnamefont
  {L.}~\bibnamefont {Huang}}, \bibinfo {author} {\bibfnamefont {C.-C.}\
  \bibnamefont {Lee}}, \bibinfo {author} {\bibfnamefont {S.-M.}\ \bibnamefont
  {Huang}}, \bibinfo {author} {\bibfnamefont {B.}~\bibnamefont {Wang}},
  \bibinfo {author} {\bibfnamefont {A.}~\bibnamefont {Bansil}}, \bibinfo
  {author} {\bibfnamefont {H.-T.}\ \bibnamefont {Jeng}}, \bibinfo {author}
  {\bibfnamefont {T.}~\bibnamefont {Neupert}}, \bibinfo {author} {\bibfnamefont
  {A.}~\bibnamefont {Kaminski}}, \bibinfo {author} {\bibfnamefont
  {H.}~\bibnamefont {Lin}}, \bibinfo {author} {\bibfnamefont {S.}~\bibnamefont
  {Jia}}, \ and\ \bibinfo {author} {\bibfnamefont {M.~Z.}\ \bibnamefont
  {Hasan}},\ }\bibfield  {title} {\enquote {\bibinfo {title} {Discovery of a
  {W}eyl fermion state with {F}ermi arcs in niobium arsenide},}\ }\href
  {\doibase 10.1038/nphys3437} {\bibfield  {journal} {\bibinfo  {journal} {Nat.
  Phys.}\ }\textbf {\bibinfo {volume} {11}},\ \bibinfo {pages} {748--754}
  (\bibinfo {year} {2015}{\natexlab{b}})}\BibitemShut {NoStop}%
\bibitem [{\citenamefont {Xu}\ \emph {et~al.}(2015{\natexlab{c}})\citenamefont
  {Xu}, \citenamefont {Belopolski}, \citenamefont {Sanchez}, \citenamefont
  {Zhang}, \citenamefont {Chang}, \citenamefont {Guo}, \citenamefont {Bian},
  \citenamefont {Yuan}, \citenamefont {Lu}, \citenamefont {Chang},
  \citenamefont {Shibayev}, \citenamefont {Prokopovych}, \citenamefont
  {Alidoust}, \citenamefont {Zheng}, \citenamefont {Lee}, \citenamefont
  {Huang}, \citenamefont {Sankar}, \citenamefont {Chou}, \citenamefont {Hsu},
  \citenamefont {Jeng}, \citenamefont {Bansil}, \citenamefont {Neupert},
  \citenamefont {Strocov}, \citenamefont {Lin}, \citenamefont {Jia},\ and\
  \citenamefont {Hasan}}]{Xu_SciAdv_2015}%
  \BibitemOpen
  \bibfield  {author} {\bibinfo {author} {\bibfnamefont {S.-Y.}\ \bibnamefont
  {Xu}}, \bibinfo {author} {\bibfnamefont {I.}~\bibnamefont {Belopolski}},
  \bibinfo {author} {\bibfnamefont {D.~S.}\ \bibnamefont {Sanchez}}, \bibinfo
  {author} {\bibfnamefont {C.}~\bibnamefont {Zhang}}, \bibinfo {author}
  {\bibfnamefont {G.}~\bibnamefont {Chang}}, \bibinfo {author} {\bibfnamefont
  {C.}~\bibnamefont {Guo}}, \bibinfo {author} {\bibfnamefont {G.}~\bibnamefont
  {Bian}}, \bibinfo {author} {\bibfnamefont {Z.}~\bibnamefont {Yuan}}, \bibinfo
  {author} {\bibfnamefont {H.}~\bibnamefont {Lu}}, \bibinfo {author}
  {\bibfnamefont {T.-R.}\ \bibnamefont {Chang}}, \bibinfo {author}
  {\bibfnamefont {P.~P.}\ \bibnamefont {Shibayev}}, \bibinfo {author}
  {\bibfnamefont {M.~L.}\ \bibnamefont {Prokopovych}}, \bibinfo {author}
  {\bibfnamefont {N.}~\bibnamefont {Alidoust}}, \bibinfo {author}
  {\bibfnamefont {H.}~\bibnamefont {Zheng}}, \bibinfo {author} {\bibfnamefont
  {C.-C.}\ \bibnamefont {Lee}}, \bibinfo {author} {\bibfnamefont {S.-M.}\
  \bibnamefont {Huang}}, \bibinfo {author} {\bibfnamefont {R.}~\bibnamefont
  {Sankar}}, \bibinfo {author} {\bibfnamefont {F.}~\bibnamefont {Chou}},
  \bibinfo {author} {\bibfnamefont {C.-H.}\ \bibnamefont {Hsu}}, \bibinfo
  {author} {\bibfnamefont {H.-T.}\ \bibnamefont {Jeng}}, \bibinfo {author}
  {\bibfnamefont {A.}~\bibnamefont {Bansil}}, \bibinfo {author} {\bibfnamefont
  {T.}~\bibnamefont {Neupert}}, \bibinfo {author} {\bibfnamefont {V.~N.}\
  \bibnamefont {Strocov}}, \bibinfo {author} {\bibfnamefont {H.}~\bibnamefont
  {Lin}}, \bibinfo {author} {\bibfnamefont {S.}~\bibnamefont {Jia}}, \ and\
  \bibinfo {author} {\bibfnamefont {M.~Z.}\ \bibnamefont {Hasan}},\ }\bibfield
  {title} {\enquote {\bibinfo {title} {Experimental discovery of a topological
  {W}eyl semimetal state in {T}a{P}},}\ }\href {\doibase
  10.1126/sciadv.1501092} {\bibfield  {journal} {\bibinfo  {journal} {Sci.
  Adv.}\ }\textbf {\bibinfo {volume} {1}},\ \bibinfo {pages} {e1501092}
  (\bibinfo {year} {2015}{\natexlab{c}})},\ \Eprint
  {http://arxiv.org/abs/arXiv:1508.03102} {arXiv:1508.03102} \BibitemShut
  {NoStop}%
\bibitem [{\citenamefont {Lee}\ \emph {et~al.}(2015)\citenamefont {Lee},
  \citenamefont {Xu}, \citenamefont {Huang}, \citenamefont {Sanchez},
  \citenamefont {Belopolski}, \citenamefont {Chang}, \citenamefont {Bian},
  \citenamefont {Alidoust}, \citenamefont {Zheng}, \citenamefont {Neupane},
  \citenamefont {Wang}, \citenamefont {Bansil}, \citenamefont {Hasan},\ and\
  \citenamefont {Lin}}]{Lee_ARXIV_2015}%
  \BibitemOpen
  \bibfield  {author} {\bibinfo {author} {\bibfnamefont {C.-C.}\ \bibnamefont
  {Lee}}, \bibinfo {author} {\bibfnamefont {S.-Y.}\ \bibnamefont {Xu}},
  \bibinfo {author} {\bibfnamefont {S.-M.}\ \bibnamefont {Huang}}, \bibinfo
  {author} {\bibfnamefont {D.~S.}\ \bibnamefont {Sanchez}}, \bibinfo {author}
  {\bibfnamefont {I.}~\bibnamefont {Belopolski}}, \bibinfo {author}
  {\bibfnamefont {G.}~\bibnamefont {Chang}}, \bibinfo {author} {\bibfnamefont
  {G.}~\bibnamefont {Bian}}, \bibinfo {author} {\bibfnamefont {N.}~\bibnamefont
  {Alidoust}}, \bibinfo {author} {\bibfnamefont {H.}~\bibnamefont {Zheng}},
  \bibinfo {author} {\bibfnamefont {M.}~\bibnamefont {Neupane}}, \bibinfo
  {author} {\bibfnamefont {B.}~\bibnamefont {Wang}}, \bibinfo {author}
  {\bibfnamefont {A.}~\bibnamefont {Bansil}}, \bibinfo {author} {\bibfnamefont
  {M.~Z.}\ \bibnamefont {Hasan}}, \ and\ \bibinfo {author} {\bibfnamefont
  {H.}~\bibnamefont {Lin}},\ }\href@noop {} {\enquote {\bibinfo {title} {Fermi
  arc topology and interconnectivity in {W}eyl fermion semimetals {T}a{A}s,
  {T}a{P}, {N}b{A}s, and {N}b{P}},}\ } (\bibinfo {year} {2015}),\ \Eprint
  {http://arxiv.org/abs/arXiv:1508.05999} {arXiv:1508.05999} \BibitemShut
  {NoStop}%
\bibitem [{\citenamefont {Sun}\ \emph {et~al.}(2015)\citenamefont {Sun},
  \citenamefont {Wu},\ and\ \citenamefont {Yan}}]{Sun_PRB_2015}%
  \BibitemOpen
  \bibfield  {author} {\bibinfo {author} {\bibfnamefont {Y.}~\bibnamefont
  {Sun}}, \bibinfo {author} {\bibfnamefont {S.-C.}\ \bibnamefont {Wu}}, \ and\
  \bibinfo {author} {\bibfnamefont {B.}~\bibnamefont {Yan}},\ }\bibfield
  {title} {\enquote {\bibinfo {title} {Topological surface states and {F}ermi
  arcs of the noncentrosymmetric {W}eyl semimetals {T}a{A}s, {T}a{P}, {N}b{A}s,
  and {N}b{P}},}\ }\href {\doibase 10.1103/PhysRevB.92.115428} {\bibfield
  {journal} {\bibinfo  {journal} {Phys. Rev. B}\ }\textbf {\bibinfo {volume}
  {92}},\ \bibinfo {pages} {115428} (\bibinfo {year} {2015})}\BibitemShut
  {NoStop}%
\bibitem [{\citenamefont {Lv}\ \emph {et~al.}(2015{\natexlab{c}})\citenamefont
  {Lv}, \citenamefont {Muff}, \citenamefont {Qian}, \citenamefont {Song},
  \citenamefont {Nie}, \citenamefont {Xu}, \citenamefont {Richard},
  \citenamefont {Matt}, \citenamefont {Plumb}, \citenamefont {Zhao},
  \citenamefont {Chen}, \citenamefont {Fang}, \citenamefont {Dai},
  \citenamefont {Dil}, \citenamefont {Mesot}, \citenamefont {Shi},
  \citenamefont {Weng},\ and\ \citenamefont {Ding}}]{Lv_PRL_2015}%
  \BibitemOpen
  \bibfield  {author} {\bibinfo {author} {\bibfnamefont {B.~Q.}\ \bibnamefont
  {Lv}}, \bibinfo {author} {\bibfnamefont {S.}~\bibnamefont {Muff}}, \bibinfo
  {author} {\bibfnamefont {T.}~\bibnamefont {Qian}}, \bibinfo {author}
  {\bibfnamefont {Z.D.}\ \bibnamefont {Song}}, \bibinfo {author} {\bibfnamefont
  {S.~M.}\ \bibnamefont {Nie}}, \bibinfo {author} {\bibfnamefont
  {N.}~\bibnamefont {Xu}}, \bibinfo {author} {\bibfnamefont {P.}~\bibnamefont
  {Richard}}, \bibinfo {author} {\bibfnamefont {C.~E.}\ \bibnamefont {Matt}},
  \bibinfo {author} {\bibfnamefont {N.~C.}\ \bibnamefont {Plumb}}, \bibinfo
  {author} {\bibfnamefont {L.~X.}\ \bibnamefont {Zhao}}, \bibinfo {author}
  {\bibfnamefont {G.~F.}\ \bibnamefont {Chen}}, \bibinfo {author}
  {\bibfnamefont {Z.}~\bibnamefont {Fang}}, \bibinfo {author} {\bibfnamefont
  {X.}~\bibnamefont {Dai}}, \bibinfo {author} {\bibfnamefont {J.~H.}\
  \bibnamefont {Dil}}, \bibinfo {author} {\bibfnamefont {J.}~\bibnamefont
  {Mesot}}, \bibinfo {author} {\bibfnamefont {M.}~\bibnamefont {Shi}}, \bibinfo
  {author} {\bibfnamefont {H.~M.}\ \bibnamefont {Weng}}, \ and\ \bibinfo
  {author} {\bibfnamefont {H.}~\bibnamefont {Ding}},\ }\bibfield  {title}
  {\enquote {\bibinfo {title} {Observation of {F}ermi-arc spin texture in
  {T}a{A}s},}\ }\href {\doibase 10.1103/PhysRevLett.115.217601} {\bibfield
  {journal} {\bibinfo  {journal} {Phys. Rev. Lett.}\ }\textbf {\bibinfo
  {volume} {115}},\ \bibinfo {pages} {217601} (\bibinfo {year}
  {2015}{\natexlab{c}})}\BibitemShut {NoStop}%
\bibitem [{\citenamefont {Zhang}\ \emph
  {et~al.}(2015{\natexlab{a}})\citenamefont {Zhang}, \citenamefont {Guo},
  \citenamefont {Lu}, \citenamefont {Zhang}, \citenamefont {Yuan},
  \citenamefont {Lin}, \citenamefont {Wang},\ and\ \citenamefont
  {Jia}}]{Zhang_PRB_2015}%
  \BibitemOpen
  \bibfield  {author} {\bibinfo {author} {\bibfnamefont {C.}~\bibnamefont
  {Zhang}}, \bibinfo {author} {\bibfnamefont {C.}~\bibnamefont {Guo}}, \bibinfo
  {author} {\bibfnamefont {H.}~\bibnamefont {Lu}}, \bibinfo {author}
  {\bibfnamefont {X.}~\bibnamefont {Zhang}}, \bibinfo {author} {\bibfnamefont
  {Z.}~\bibnamefont {Yuan}}, \bibinfo {author} {\bibfnamefont {Z.}~\bibnamefont
  {Lin}}, \bibinfo {author} {\bibfnamefont {J.}~\bibnamefont {Wang}}, \ and\
  \bibinfo {author} {\bibfnamefont {S.}~\bibnamefont {Jia}},\ }\bibfield
  {title} {\enquote {\bibinfo {title} {Large magnetoresistance over an extended
  temperature regime in monophosphides of tantalum and niobium},}\ }\href
  {\doibase 10.1103/PhysRevB.92.041203} {\bibfield  {journal} {\bibinfo
  {journal} {Phys. Rev. B}\ }\textbf {\bibinfo {volume} {92}},\ \bibinfo
  {pages} {041203} (\bibinfo {year} {2015}{\natexlab{a}})}\BibitemShut
  {NoStop}%
\bibitem [{\citenamefont {Huang}\ \emph
  {et~al.}(2015{\natexlab{b}})\citenamefont {Huang}, \citenamefont {Zhao},
  \citenamefont {Long}, \citenamefont {Wang}, \citenamefont {Chen},
  \citenamefont {Yang}, \citenamefont {Liang}, \citenamefont {Xue},
  \citenamefont {Weng}, \citenamefont {Fang}, \citenamefont {Dai},\ and\
  \citenamefont {Chen}}]{Huang_PRX_2015}%
  \BibitemOpen
  \bibfield  {author} {\bibinfo {author} {\bibfnamefont {X.}~\bibnamefont
  {Huang}}, \bibinfo {author} {\bibfnamefont {L.}~\bibnamefont {Zhao}},
  \bibinfo {author} {\bibfnamefont {Y.}~\bibnamefont {Long}}, \bibinfo {author}
  {\bibfnamefont {P.}~\bibnamefont {Wang}}, \bibinfo {author} {\bibfnamefont
  {D.}~\bibnamefont {Chen}}, \bibinfo {author} {\bibfnamefont {Z.}~\bibnamefont
  {Yang}}, \bibinfo {author} {\bibfnamefont {H.}~\bibnamefont {Liang}},
  \bibinfo {author} {\bibfnamefont {M.}~\bibnamefont {Xue}}, \bibinfo {author}
  {\bibfnamefont {H.}~\bibnamefont {Weng}}, \bibinfo {author} {\bibfnamefont
  {Z.}~\bibnamefont {Fang}}, \bibinfo {author} {\bibfnamefont {X.}~\bibnamefont
  {Dai}}, \ and\ \bibinfo {author} {\bibfnamefont {G.}~\bibnamefont {Chen}},\
  }\bibfield  {title} {\enquote {\bibinfo {title} {Observation of the
  chiral-anomaly-induced negative magnetoresistance in 3{D} {W}eyl semimetal
  {T}a{A}s},}\ }\href {\doibase 10.1103/PhysRevX.5.031023} {\bibfield
  {journal} {\bibinfo  {journal} {Phys. Rev. X}\ }\textbf {\bibinfo {volume}
  {5}},\ \bibinfo {pages} {031023} (\bibinfo {year}
  {2015}{\natexlab{b}})}\BibitemShut {NoStop}%
\bibitem [{\citenamefont {Shekhar}\ \emph
  {et~al.}(2015{\natexlab{a}})\citenamefont {Shekhar}, \citenamefont {Nayak},
  \citenamefont {Sun}, \citenamefont {Schmidt}, \citenamefont {Nicklas},
  \citenamefont {Leermakers}, \citenamefont {Zeitler}, \citenamefont
  {Skourski}, \citenamefont {Wosnitza}, \citenamefont {Liu}, \citenamefont
  {Chen}, \citenamefont {Schnelle}, \citenamefont {Borrmann}, \citenamefont
  {Grin}, \citenamefont {Felser},\ and\ \citenamefont
  {Yan}}]{Shekhar_NatPhys_2015}%
  \BibitemOpen
  \bibfield  {author} {\bibinfo {author} {\bibfnamefont {C.}~\bibnamefont
  {Shekhar}}, \bibinfo {author} {\bibfnamefont {A.~K.}\ \bibnamefont {Nayak}},
  \bibinfo {author} {\bibfnamefont {Y.}~\bibnamefont {Sun}}, \bibinfo {author}
  {\bibfnamefont {M.}~\bibnamefont {Schmidt}}, \bibinfo {author} {\bibfnamefont
  {M.}~\bibnamefont {Nicklas}}, \bibinfo {author} {\bibfnamefont
  {I.}~\bibnamefont {Leermakers}}, \bibinfo {author} {\bibfnamefont
  {U.}~\bibnamefont {Zeitler}}, \bibinfo {author} {\bibfnamefont
  {Y.}~\bibnamefont {Skourski}}, \bibinfo {author} {\bibfnamefont
  {J.}~\bibnamefont {Wosnitza}}, \bibinfo {author} {\bibfnamefont
  {Z.}~\bibnamefont {Liu}}, \bibinfo {author} {\bibfnamefont {Y.}~\bibnamefont
  {Chen}}, \bibinfo {author} {\bibfnamefont {W.}~\bibnamefont {Schnelle}},
  \bibinfo {author} {\bibfnamefont {H.}~\bibnamefont {Borrmann}}, \bibinfo
  {author} {\bibfnamefont {Y.}~\bibnamefont {Grin}}, \bibinfo {author}
  {\bibfnamefont {C.}~\bibnamefont {Felser}}, \ and\ \bibinfo {author}
  {\bibfnamefont {B.}~\bibnamefont {Yan}},\ }\bibfield  {title} {\enquote
  {\bibinfo {title} {Extremely large magnetoresistance and ultrahigh mobility
  in the topological {W}eyl semimetal candidate {N}b{P}},}\ }\href {\doibase
  10.1038/nphys3372} {\bibfield  {journal} {\bibinfo  {journal} {Nat. Phys.}\
  }\textbf {\bibinfo {volume} {11}},\ \bibinfo {pages} {645--649} (\bibinfo
  {year} {2015}{\natexlab{a}})}\BibitemShut {NoStop}%
\bibitem [{\citenamefont {Ghimire}\ \emph {et~al.}(2015)\citenamefont
  {Ghimire}, \citenamefont {Luo}, \citenamefont {Neupane}, \citenamefont
  {Williams}, \citenamefont {Bauer},\ and\ \citenamefont
  {Ronning}}]{Ghimire_JPCM_2015}%
  \BibitemOpen
  \bibfield  {author} {\bibinfo {author} {\bibfnamefont {N.~J.}\ \bibnamefont
  {Ghimire}}, \bibinfo {author} {\bibfnamefont {Y.}~\bibnamefont {Luo}},
  \bibinfo {author} {\bibfnamefont {M.}~\bibnamefont {Neupane}}, \bibinfo
  {author} {\bibfnamefont {D.~J.}\ \bibnamefont {Williams}}, \bibinfo {author}
  {\bibfnamefont {E.~D.}\ \bibnamefont {Bauer}}, \ and\ \bibinfo {author}
  {\bibfnamefont {F.}~\bibnamefont {Ronning}},\ }\bibfield  {title} {\enquote
  {\bibinfo {title} {Magnetotransport of single crystalline {N}b{A}s},}\ }\href
  {\doibase 10.1088/0953-8984/27/15/152201} {\bibfield  {journal} {\bibinfo
  {journal} {J. Phys. Condens. Matter}\ }\textbf {\bibinfo {volume} {27}},\
  \bibinfo {pages} {152201} (\bibinfo {year} {2015})}\BibitemShut {NoStop}%
\bibitem [{\citenamefont {Zhang}\ \emph
  {et~al.}(2015{\natexlab{b}})\citenamefont {Zhang}, \citenamefont {Yuan},
  \citenamefont {Xu}, \citenamefont {Lin}, \citenamefont {Tong}, \citenamefont
  {Hasan}, \citenamefont {Wang}, \citenamefont {Zang},\ and\ \citenamefont
  {Jia}}]{Zhang_ARXIV_2015a}%
  \BibitemOpen
  \bibfield  {author} {\bibinfo {author} {\bibfnamefont {C.}~\bibnamefont
  {Zhang}}, \bibinfo {author} {\bibfnamefont {Z.}~\bibnamefont {Yuan}},
  \bibinfo {author} {\bibfnamefont {S.}~\bibnamefont {Xu}}, \bibinfo {author}
  {\bibfnamefont {Z.}~\bibnamefont {Lin}}, \bibinfo {author} {\bibfnamefont
  {B.}~\bibnamefont {Tong}}, \bibinfo {author} {\bibfnamefont {M.~Z.}\
  \bibnamefont {Hasan}}, \bibinfo {author} {\bibfnamefont {J.}~\bibnamefont
  {Wang}}, \bibinfo {author} {\bibfnamefont {C.}~\bibnamefont {Zang}}, \ and\
  \bibinfo {author} {\bibfnamefont {S.}~\bibnamefont {Jia}},\ }\href@noop {}
  {\enquote {\bibinfo {title} {Tantalum monoarsenide: an exotic compensated
  semimetal},}\ } (\bibinfo {year} {2015}{\natexlab{b}}),\ \Eprint
  {http://arxiv.org/abs/arXiv:1502.00251} {arXiv:1502.00251} \BibitemShut
  {NoStop}%
\bibitem [{\citenamefont {Shekhar}\ \emph
  {et~al.}(2015{\natexlab{b}})\citenamefont {Shekhar}, \citenamefont {Arnold},
  \citenamefont {Wu}, \citenamefont {Sun}, \citenamefont {Schmidt},
  \citenamefont {Kumar}, \citenamefont {Grushin}, \citenamefont {Bardarson},
  \citenamefont {{Donizeth dos Reis}}, \citenamefont {Naumann}, \citenamefont
  {Baenitz}, \citenamefont {Borrmann}, \citenamefont {Nicklas}, \citenamefont
  {Hassinger}, \citenamefont {Felser},\ and\ \citenamefont
  {Yan}}]{Shekhar_ARXIV_2015}%
  \BibitemOpen
  \bibfield  {author} {\bibinfo {author} {\bibfnamefont {C.}~\bibnamefont
  {Shekhar}}, \bibinfo {author} {\bibfnamefont {F.}~\bibnamefont {Arnold}},
  \bibinfo {author} {\bibfnamefont {S.-C.}\ \bibnamefont {Wu}}, \bibinfo
  {author} {\bibfnamefont {Y.}~\bibnamefont {Sun}}, \bibinfo {author}
  {\bibfnamefont {M.}~\bibnamefont {Schmidt}}, \bibinfo {author} {\bibfnamefont
  {N.}~\bibnamefont {Kumar}}, \bibinfo {author} {\bibfnamefont {A.~G.}\
  \bibnamefont {Grushin}}, \bibinfo {author} {\bibfnamefont {J.~H.}\
  \bibnamefont {Bardarson}}, \bibinfo {author} {\bibfnamefont {R.}~\bibnamefont
  {{Donizeth dos Reis}}}, \bibinfo {author} {\bibfnamefont {M.}~\bibnamefont
  {Naumann}}, \bibinfo {author} {\bibfnamefont {M.}~\bibnamefont {Baenitz}},
  \bibinfo {author} {\bibfnamefont {H.}~\bibnamefont {Borrmann}}, \bibinfo
  {author} {\bibfnamefont {M.}~\bibnamefont {Nicklas}}, \bibinfo {author}
  {\bibfnamefont {E.}~\bibnamefont {Hassinger}}, \bibinfo {author}
  {\bibfnamefont {C.}~\bibnamefont {Felser}}, \ and\ \bibinfo {author}
  {\bibfnamefont {B.}~\bibnamefont {Yan}},\ }\href@noop {} {\enquote {\bibinfo
  {title} {Large and unsaturated negative magnetoresistance induced by the
  chiral anomaly in the {W}eyl semimetal {T}a{P}},}\ } (\bibinfo {year}
  {2015}{\natexlab{b}}),\ \Eprint {http://arxiv.org/abs/arXiv:1506.06577}
  {arXiv:1506.06577} \BibitemShut {NoStop}%
\bibitem [{\citenamefont {Wang}\ \emph {et~al.}(2015)\citenamefont {Wang},
  \citenamefont {Zheng}, \citenamefont {Shen}, \citenamefont {Zhou},
  \citenamefont {Yang}, \citenamefont {Li}, \citenamefont {Feng},\ and\
  \citenamefont {Xu}}]{Wang_ARXIV_2015}%
  \BibitemOpen
  \bibfield  {author} {\bibinfo {author} {\bibfnamefont {Z.}~\bibnamefont
  {Wang}}, \bibinfo {author} {\bibfnamefont {Y.}~\bibnamefont {Zheng}},
  \bibinfo {author} {\bibfnamefont {Z.}~\bibnamefont {Shen}}, \bibinfo {author}
  {\bibfnamefont {Y.}~\bibnamefont {Zhou}}, \bibinfo {author} {\bibfnamefont
  {Z.}~\bibnamefont {Yang}}, \bibinfo {author} {\bibfnamefont {Y.}~\bibnamefont
  {Li}}, \bibinfo {author} {\bibfnamefont {C.}~\bibnamefont {Feng}}, \ and\
  \bibinfo {author} {\bibfnamefont {Z.-A.}\ \bibnamefont {Xu}},\ }\href@noop {}
  {\enquote {\bibinfo {title} {Helicity protected ultrahigh mobility weyl
  fermions in {N}b{P}},}\ } (\bibinfo {year} {2015}),\ \Eprint
  {http://arxiv.org/abs/arXiv:1506.00924} {arXiv:1506.00924} \BibitemShut
  {NoStop}%
\bibitem [{\citenamefont {Zhang}\ \emph
  {et~al.}(2015{\natexlab{c}})\citenamefont {Zhang}, \citenamefont {Lin},
  \citenamefont {Guo}, \citenamefont {Xu}, \citenamefont {Lee}, \citenamefont
  {Lu}, \citenamefont {Huang}, \citenamefont {Chang}, \citenamefont {Hsu},
  \citenamefont {Lin}, \citenamefont {Li}, \citenamefont {Zhang}, \citenamefont
  {Neupert}, \citenamefont {Hasan}, \citenamefont {Wang},\ and\ \citenamefont
  {Jia}}]{Zhang_ARXIV_2015}%
  \BibitemOpen
  \bibfield  {author} {\bibinfo {author} {\bibfnamefont {C.}~\bibnamefont
  {Zhang}}, \bibinfo {author} {\bibfnamefont {Z.}~\bibnamefont {Lin}}, \bibinfo
  {author} {\bibfnamefont {C.}~\bibnamefont {Guo}}, \bibinfo {author}
  {\bibfnamefont {S.-Y.}\ \bibnamefont {Xu}}, \bibinfo {author} {\bibfnamefont
  {C.-C.}\ \bibnamefont {Lee}}, \bibinfo {author} {\bibfnamefont
  {H.}~\bibnamefont {Lu}}, \bibinfo {author} {\bibfnamefont {S.-M.}\
  \bibnamefont {Huang}}, \bibinfo {author} {\bibfnamefont {G.}~\bibnamefont
  {Chang}}, \bibinfo {author} {\bibfnamefont {C.-H.}\ \bibnamefont {Hsu}},
  \bibinfo {author} {\bibfnamefont {H.}~\bibnamefont {Lin}}, \bibinfo {author}
  {\bibfnamefont {L.}~\bibnamefont {Li}}, \bibinfo {author} {\bibfnamefont
  {C.}~\bibnamefont {Zhang}}, \bibinfo {author} {\bibfnamefont
  {T.}~\bibnamefont {Neupert}}, \bibinfo {author} {\bibfnamefont {M.~Z.}\
  \bibnamefont {Hasan}}, \bibinfo {author} {\bibfnamefont {J.}~\bibnamefont
  {Wang}}, \ and\ \bibinfo {author} {\bibfnamefont {S.}~\bibnamefont {Jia}},\
  }\href@noop {} {\enquote {\bibinfo {title} {Quantum phase transitions in
  {W}eyl semimetal tantalum monophosphide},}\ } (\bibinfo {year}
  {2015}{\natexlab{c}}),\ \Eprint {http://arxiv.org/abs/arXiv:1507.06301}
  {arXiv:1507.06301} \BibitemShut {NoStop}%
\bibitem [{\citenamefont {Liu}\ \emph {et~al.}(2015{\natexlab{a}})\citenamefont
  {Liu}, \citenamefont {Richard}, \citenamefont {Song}, \citenamefont {Zhao},
  \citenamefont {Fang}, \citenamefont {Chen},\ and\ \citenamefont
  {Ding}}]{Liu_PRB_2015}%
  \BibitemOpen
  \bibfield  {author} {\bibinfo {author} {\bibfnamefont {H.~W.}\ \bibnamefont
  {Liu}}, \bibinfo {author} {\bibfnamefont {P.}~\bibnamefont {Richard}},
  \bibinfo {author} {\bibfnamefont {Z.~D.}\ \bibnamefont {Song}}, \bibinfo
  {author} {\bibfnamefont {L.~X.}\ \bibnamefont {Zhao}}, \bibinfo {author}
  {\bibfnamefont {Z.}~\bibnamefont {Fang}}, \bibinfo {author} {\bibfnamefont
  {G.-F.}\ \bibnamefont {Chen}}, \ and\ \bibinfo {author} {\bibfnamefont
  {H.}~\bibnamefont {Ding}},\ }\bibfield  {title} {\enquote {\bibinfo {title}
  {Raman study of lattice dynamics in the {W}eyl semimetal {T}a{A}s},}\ }\href
  {\doibase 10.1103/PhysRevB.92.064302} {\bibfield  {journal} {\bibinfo
  {journal} {Phys. Rev. B}\ }\textbf {\bibinfo {volume} {92}},\ \bibinfo
  {pages} {064302} (\bibinfo {year} {2015}{\natexlab{a}})}\BibitemShut
  {NoStop}%
\bibitem [{\citenamefont {Liu}\ \emph {et~al.}(2015{\natexlab{b}})\citenamefont
  {Liu}, \citenamefont {Li}, \citenamefont {Guo}, \citenamefont {Chen},
  \citenamefont {Fe}, \citenamefont {Liu}, \citenamefont {Prucnal},
  \citenamefont {Helm},\ and\ \citenamefont {Zhou}}]{Liu_ARXIV_2015}%
  \BibitemOpen
  \bibfield  {author} {\bibinfo {author} {\bibfnamefont {Y.}~\bibnamefont
  {Liu}}, \bibinfo {author} {\bibfnamefont {Z.}~\bibnamefont {Li}}, \bibinfo
  {author} {\bibfnamefont {L.}~\bibnamefont {Guo}}, \bibinfo {author}
  {\bibfnamefont {X.}~\bibnamefont {Chen}}, \bibinfo {author} {\bibfnamefont
  {Y.}~\bibnamefont {Fe}}, \bibinfo {author} {\bibfnamefont {F.}~\bibnamefont
  {Liu}}, \bibinfo {author} {\bibfnamefont {S.}~\bibnamefont {Prucnal}},
  \bibinfo {author} {\bibfnamefont {M.}~\bibnamefont {Helm}}, \ and\ \bibinfo
  {author} {\bibfnamefont {S.}~\bibnamefont {Zhou}},\ }\href@noop {} {\enquote
  {\bibinfo {title} {Intrinsic diamagnetism in the {W}eyl semimetal
  {T}a{A}s},}\ } (\bibinfo {year} {2015}{\natexlab{b}}),\ \Eprint
  {http://arxiv.org/abs/arXiv:1510.08497} {arXiv:1510.08497} \BibitemShut
  {NoStop}%
\bibitem [{\citenamefont {Zhang}\ \emph
  {et~al.}(2015{\natexlab{d}})\citenamefont {Zhang}, \citenamefont {Liu},
  \citenamefont {Dong}, \citenamefont {Xu}, \citenamefont {Li}, \citenamefont
  {Yang},\ and\ \citenamefont {Li}}]{Zhang_CPL_2015}%
  \BibitemOpen
  \bibfield  {author} {\bibinfo {author} {\bibfnamefont {J.}~\bibnamefont
  {Zhang}}, \bibinfo {author} {\bibfnamefont {F.-L.}\ \bibnamefont {Liu}},
  \bibinfo {author} {\bibfnamefont {J.-K.}\ \bibnamefont {Dong}}, \bibinfo
  {author} {\bibfnamefont {Y.}~\bibnamefont {Xu}}, \bibinfo {author}
  {\bibfnamefont {N.-N.}\ \bibnamefont {Li}}, \bibinfo {author} {\bibfnamefont
  {W.-G.}\ \bibnamefont {Yang}}, \ and\ \bibinfo {author} {\bibfnamefont
  {S.-Y.}\ \bibnamefont {Li}},\ }\bibfield  {title} {\enquote {\bibinfo {title}
  {Structural and transport properties of the {W}eyl semimetal {N}b{A}s at high
  pressure},}\ }\href {\doibase 10.1088/0256-307X/32/9/097102} {\bibfield
  {journal} {\bibinfo  {journal} {Chin. Phys. Lett.}\ }\textbf {\bibinfo
  {volume} {32}},\ \bibinfo {pages} {097102} (\bibinfo {year}
  {2015}{\natexlab{d}})}\BibitemShut {NoStop}%
\bibitem [{\citenamefont {Luo}\ \emph {et~al.}(2015)\citenamefont {Luo},
  \citenamefont {Ghimire}, \citenamefont {Bauer}, \citenamefont {Thompson},\
  and\ \citenamefont {Ronning}}]{Luo_ARXIV_2015}%
  \BibitemOpen
  \bibfield  {author} {\bibinfo {author} {\bibfnamefont {Y.}~\bibnamefont
  {Luo}}, \bibinfo {author} {\bibfnamefont {N.~J.}\ \bibnamefont {Ghimire}},
  \bibinfo {author} {\bibfnamefont {E.~D.}\ \bibnamefont {Bauer}}, \bibinfo
  {author} {\bibfnamefont {J.~D.}\ \bibnamefont {Thompson}}, \ and\ \bibinfo
  {author} {\bibfnamefont {F.}~\bibnamefont {Ronning}},\ }\href@noop {}
  {\enquote {\bibinfo {title} {"{H}ard" crystalline lattice in the {W}eyl
  semimetal {N}b{A}s},}\ } (\bibinfo {year} {2015}),\ \Eprint
  {http://arxiv.org/abs/arXiv:1510.08538} {arXiv:1510.08538} \BibitemShut
  {NoStop}%
\bibitem [{\citenamefont {Zheng}\ \emph {et~al.}(2015)\citenamefont {Zheng},
  \citenamefont {Xu}, \citenamefont {Bian}, \citenamefont {Guo}, \citenamefont
  {Chang}, \citenamefont {Sanchez}, \citenamefont {Belopolski}, \citenamefont
  {Lee}, \citenamefont {Huang}, \citenamefont {Zhang}, \citenamefont {Sankar},
  \citenamefont {Alidoust}, \citenamefont {Chang}, \citenamefont {Wu},
  \citenamefont {Neupert}, \citenamefont {Chou}, \citenamefont {Jeng},
  \citenamefont {Yao}, \citenamefont {Bansil}, \citenamefont {Jia},
  \citenamefont {Lin},\ and\ \citenamefont {Hasan}}]{Zheng_ARXIV_2015}%
  \BibitemOpen
  \bibfield  {author} {\bibinfo {author} {\bibfnamefont {H.}~\bibnamefont
  {Zheng}}, \bibinfo {author} {\bibfnamefont {S.-Y.}\ \bibnamefont {Xu}},
  \bibinfo {author} {\bibfnamefont {G.}~\bibnamefont {Bian}}, \bibinfo {author}
  {\bibfnamefont {C.}~\bibnamefont {Guo}}, \bibinfo {author} {\bibfnamefont
  {G.}~\bibnamefont {Chang}}, \bibinfo {author} {\bibfnamefont {D.~S.}\
  \bibnamefont {Sanchez}}, \bibinfo {author} {\bibfnamefont {I.}~\bibnamefont
  {Belopolski}}, \bibinfo {author} {\bibfnamefont {C.-C.}\ \bibnamefont {Lee}},
  \bibinfo {author} {\bibfnamefont {S.-M.}\ \bibnamefont {Huang}}, \bibinfo
  {author} {\bibfnamefont {X.}~\bibnamefont {Zhang}}, \bibinfo {author}
  {\bibfnamefont {R.}~\bibnamefont {Sankar}}, \bibinfo {author} {\bibfnamefont
  {N.}~\bibnamefont {Alidoust}}, \bibinfo {author} {\bibfnamefont {T.-R.}\
  \bibnamefont {Chang}}, \bibinfo {author} {\bibfnamefont {F.}~\bibnamefont
  {Wu}}, \bibinfo {author} {\bibfnamefont {T.}~\bibnamefont {Neupert}},
  \bibinfo {author} {\bibfnamefont {F.}~\bibnamefont {Chou}}, \bibinfo {author}
  {\bibfnamefont {H.-T.}\ \bibnamefont {Jeng}}, \bibinfo {author}
  {\bibfnamefont {N.}~\bibnamefont {Yao}}, \bibinfo {author} {\bibfnamefont
  {A.}~\bibnamefont {Bansil}}, \bibinfo {author} {\bibfnamefont
  {S.}~\bibnamefont {Jia}}, \bibinfo {author} {\bibfnamefont {H.}~\bibnamefont
  {Lin}}, \ and\ \bibinfo {author} {\bibfnamefont {M.~Z.}\ \bibnamefont
  {Hasan}},\ }\href@noop {} {\enquote {\bibinfo {title} {Atomic scale
  visualization of quantum interference on a {W}eyl semimetal surface by
  scanning tunneling microscopy spectroscopy},}\ } (\bibinfo {year} {2015}),\
  \Eprint {http://arxiv.org/abs/arXiv:1511.02216} {arXiv:1511.02216}
  \BibitemShut {NoStop}%
\bibitem [{\citenamefont {Chang}\ \emph {et~al.}(2015)\citenamefont {Chang},
  \citenamefont {Xu}, \citenamefont {Zheng}, \citenamefont {Lee}, \citenamefont
  {Huang}, \citenamefont {Belopolski}, \citenamefont {Sanchez}, \citenamefont
  {Bian}, \citenamefont {Alidoust}, \citenamefont {Chang}, \citenamefont {Hsu},
  \citenamefont {Jeng}, \citenamefont {Bansil}, \citenamefont {Lin},\ and\
  \citenamefont {Hasan}}]{Chang_ARXIV_2015}%
  \BibitemOpen
  \bibfield  {author} {\bibinfo {author} {\bibfnamefont {G.}~\bibnamefont
  {Chang}}, \bibinfo {author} {\bibfnamefont {S.-Y.}\ \bibnamefont {Xu}},
  \bibinfo {author} {\bibfnamefont {H.}~\bibnamefont {Zheng}}, \bibinfo
  {author} {\bibfnamefont {C.-C.}\ \bibnamefont {Lee}}, \bibinfo {author}
  {\bibfnamefont {S.-M.}\ \bibnamefont {Huang}}, \bibinfo {author}
  {\bibfnamefont {I.}~\bibnamefont {Belopolski}}, \bibinfo {author}
  {\bibfnamefont {D.~S.}\ \bibnamefont {Sanchez}}, \bibinfo {author}
  {\bibfnamefont {G.}~\bibnamefont {Bian}}, \bibinfo {author} {\bibfnamefont
  {N.}~\bibnamefont {Alidoust}}, \bibinfo {author} {\bibfnamefont {T.-R.}\
  \bibnamefont {Chang}}, \bibinfo {author} {\bibfnamefont {C.-H.}\ \bibnamefont
  {Hsu}}, \bibinfo {author} {\bibfnamefont {H.-T.}\ \bibnamefont {Jeng}},
  \bibinfo {author} {\bibfnamefont {A.}~\bibnamefont {Bansil}}, \bibinfo
  {author} {\bibfnamefont {H.}~\bibnamefont {Lin}}, \ and\ \bibinfo {author}
  {\bibfnamefont {M.~Z.}\ \bibnamefont {Hasan}},\ }\href@noop {} {\enquote
  {\bibinfo {title} {Quasi-particle interferences of the {W}eyl semimetals
  {T}a{A}s and {N}b{P}},}\ } (\bibinfo {year} {2015}),\ \Eprint
  {http://arxiv.org/abs/arXiv:1511.02827} {arXiv:1511.02827} \BibitemShut
  {NoStop}%
\bibitem [{\citenamefont {Sch\"{o}nberg}(1954)}]{Schonberg_ACS_1954}%
  \BibitemOpen
  \bibfield  {author} {\bibinfo {author} {\bibfnamefont {N.}~\bibnamefont
  {Sch\"{o}nberg}},\ }\bibfield  {title} {\enquote {\bibinfo {title} {An x-ray
  investigation of transition metal phosphides},}\ }\href@noop {} {\bibfield
  {journal} {\bibinfo  {journal} {Acta Chem. Scand.}\ }\textbf {\bibinfo
  {volume} {8}},\ \bibinfo {pages} {226--239} (\bibinfo {year}
  {1954})}\BibitemShut {NoStop}%
\bibitem [{\citenamefont {Boller}\ and\ \citenamefont
  {Parth\'{e}}(1963)}]{Boller_ActaCryst_1963}%
  \BibitemOpen
  \bibfield  {author} {\bibinfo {author} {\bibfnamefont {H.}~\bibnamefont
  {Boller}}\ and\ \bibinfo {author} {\bibfnamefont {E.}~\bibnamefont
  {Parth\'{e}}},\ }\bibfield  {title} {\enquote {\bibinfo {title} {The
  transposition structure of {N}b{A}s and of similar monophosphides and
  arsenides of niobium and tantalum},}\ }\href@noop {} {\bibfield  {journal}
  {\bibinfo  {journal} {Acta. Cryst.}\ }\textbf {\bibinfo {volume} {16}},\
  \bibinfo {pages} {1095--1101} (\bibinfo {year} {1963})}\BibitemShut {NoStop}%
\bibitem [{\citenamefont {Furuseth}\ and\ \citenamefont
  {Kjekshus}(1964{\natexlab{a}})}]{Furuseth_ActaCryst_1964}%
  \BibitemOpen
  \bibfield  {author} {\bibinfo {author} {\bibfnamefont {S.}~\bibnamefont
  {Furuseth}}\ and\ \bibinfo {author} {\bibfnamefont {A.}~\bibnamefont
  {Kjekshus}},\ }\bibfield  {title} {\enquote {\bibinfo {title} {The crystal
  structure of {N}b{A}s (comments)},}\ }\href@noop {} {\bibfield  {journal}
  {\bibinfo  {journal} {Acta. Cryst.}\ }\textbf {\bibinfo {volume} {17}},\
  \bibinfo {pages} {1077--1078} (\bibinfo {year}
  {1964}{\natexlab{a}})}\BibitemShut {NoStop}%
\bibitem [{\citenamefont {Furuseth}\ and\ \citenamefont
  {Kjekshus}(1964{\natexlab{b}})}]{Furuseth_Nature_1964}%
  \BibitemOpen
  \bibfield  {author} {\bibinfo {author} {\bibfnamefont {S.}~\bibnamefont
  {Furuseth}}\ and\ \bibinfo {author} {\bibfnamefont {A.}~\bibnamefont
  {Kjekshus}},\ }\bibfield  {title} {\enquote {\bibinfo {title} {Arsenides and
  antimonides of niobium},}\ }\href@noop {} {\bibfield  {journal} {\bibinfo
  {journal} {Nature}\ }\textbf {\bibinfo {volume} {203}},\ \bibinfo {pages}
  {1512} (\bibinfo {year} {1964}{\natexlab{b}})}\BibitemShut {NoStop}%
\bibitem [{\citenamefont {Saini}\ \emph {et~al.}(1964)\citenamefont {Saini},
  \citenamefont {Calvert},\ and\ \citenamefont {Taylor}}]{Saini_CanJChem_1964}%
  \BibitemOpen
  \bibfield  {author} {\bibinfo {author} {\bibfnamefont {G.~S.}\ \bibnamefont
  {Saini}}, \bibinfo {author} {\bibfnamefont {L.~D.}\ \bibnamefont {Calvert}},
  \ and\ \bibinfo {author} {\bibfnamefont {J.~B.}\ \bibnamefont {Taylor}},\
  }\bibfield  {title} {\enquote {\bibinfo {title} {Preparation and
  characterization of crystals of {MX}- and {MX}$_2$-type arsenides of niobium
  and tantalum},}\ }\href@noop {} {\bibfield  {journal} {\bibinfo  {journal}
  {Can. J. Chem.}\ }\textbf {\bibinfo {volume} {42}},\ \bibinfo {pages}
  {630--634} (\bibinfo {year} {1964})}\BibitemShut {NoStop}%
\bibitem [{\citenamefont {Furuseth}\ and\ \citenamefont
  {Kjekshus}(1964{\natexlab{c}})}]{Furuseth_ActaChemScand_1964}%
  \BibitemOpen
  \bibfield  {author} {\bibinfo {author} {\bibfnamefont {S.}~\bibnamefont
  {Furuseth}}\ and\ \bibinfo {author} {\bibfnamefont {A.}~\bibnamefont
  {Kjekshus}},\ }\bibfield  {title} {\enquote {\bibinfo {title} {On the
  arsenides and antimonides of niobium},}\ }\href@noop {} {\bibfield  {journal}
  {\bibinfo  {journal} {Acta Chem. Scand.}\ }\textbf {\bibinfo {volume} {18}},\
  \bibinfo {pages} {1180--1195} (\bibinfo {year}
  {1964}{\natexlab{c}})}\BibitemShut {NoStop}%
\bibitem [{\citenamefont {Furuseth}\ \emph {et~al.}(1965)\citenamefont
  {Furuseth}, \citenamefont {Selte},\ and\ \citenamefont
  {Kjekshus}}]{Furuseth_ActaChemScand_1965}%
  \BibitemOpen
  \bibfield  {author} {\bibinfo {author} {\bibfnamefont {S.}~\bibnamefont
  {Furuseth}}, \bibinfo {author} {\bibfnamefont {K.}~\bibnamefont {Selte}}, \
  and\ \bibinfo {author} {\bibfnamefont {A.}~\bibnamefont {Kjekshus}},\
  }\bibfield  {title} {\enquote {\bibinfo {title} {On the arsenides and
  antimonides of tantalum},}\ }\href@noop {} {\bibfield  {journal} {\bibinfo
  {journal} {Acta Chem. Scand.}\ }\textbf {\bibinfo {volume} {19}},\ \bibinfo
  {pages} {95--106} (\bibinfo {year} {1965})}\BibitemShut {NoStop}%
\bibitem [{\citenamefont {Rundqvist}(1966)}]{Rundqvist_Nature_1966}%
  \BibitemOpen
  \bibfield  {author} {\bibinfo {author} {\bibfnamefont {S.}~\bibnamefont
  {Rundqvist}},\ }\bibfield  {title} {\enquote {\bibinfo {title} {New
  metal-rich phosphides of niobium, tantalum and tungsten},}\ }\href@noop {}
  {\bibfield  {journal} {\bibinfo  {journal} {Nature}\ }\textbf {\bibinfo
  {volume} {211}},\ \bibinfo {pages} {847--848} (\bibinfo {year}
  {1966})}\BibitemShut {NoStop}%
\bibitem [{\citenamefont {Murray}\ \emph {et~al.}(1976)\citenamefont {Murray},
  \citenamefont {Taylor}, \citenamefont {Calvert}, \citenamefont {Wang},
  \citenamefont {Gabe},\ and\ \citenamefont {Despault}}]{Murray_JLess_1976}%
  \BibitemOpen
  \bibfield  {author} {\bibinfo {author} {\bibfnamefont {J.~J.}\ \bibnamefont
  {Murray}}, \bibinfo {author} {\bibfnamefont {J.~B.}\ \bibnamefont {Taylor}},
  \bibinfo {author} {\bibfnamefont {L.~D.}\ \bibnamefont {Calvert}}, \bibinfo
  {author} {\bibfnamefont {Y.}~\bibnamefont {Wang}}, \bibinfo {author}
  {\bibfnamefont {E.~J.}\ \bibnamefont {Gabe}}, \ and\ \bibinfo {author}
  {\bibfnamefont {J.~G.}\ \bibnamefont {Despault}},\ }\bibfield  {title}
  {\enquote {\bibinfo {title} {Phase relations and thermodynamics of refractory
  metal pnictides: the metal-rich tantalum arsenides},}\ }\href@noop {}
  {\bibfield  {journal} {\bibinfo  {journal} {J. Less-Common Met.}\ }\textbf
  {\bibinfo {volume} {46}},\ \bibinfo {pages} {311--320} (\bibinfo {year}
  {1976})}\BibitemShut {NoStop}%
\bibitem [{\citenamefont {Willerstr\"{o}m}(1984)}]{Willerstrom_JLess_1984}%
  \BibitemOpen
  \bibfield  {author} {\bibinfo {author} {\bibfnamefont {J.-O.}\ \bibnamefont
  {Willerstr\"{o}m}},\ }\bibfield  {title} {\enquote {\bibinfo {title}
  {Stacking disorder in {N}b{P}, {T}a{P}, {N}b{A}s and {T}a{A}s},}\ }\href@noop
  {} {\bibfield  {journal} {\bibinfo  {journal} {J. Less-Common Met.}\ }\textbf
  {\bibinfo {volume} {99}},\ \bibinfo {pages} {273--283} (\bibinfo {year}
  {1984})}\BibitemShut {NoStop}%
\bibitem [{\citenamefont {Xu}\ \emph {et~al.}(1996)\citenamefont {Xu},
  \citenamefont {Greenblatt}, \citenamefont {Emge}, \citenamefont {H\"{o}hn},
  \citenamefont {Hughbanks},\ and\ \citenamefont {Tian}}]{Xu_InorgChem_1996}%
  \BibitemOpen
  \bibfield  {author} {\bibinfo {author} {\bibfnamefont {J.}~\bibnamefont
  {Xu}}, \bibinfo {author} {\bibfnamefont {M.}~\bibnamefont {Greenblatt}},
  \bibinfo {author} {\bibfnamefont {T.}~\bibnamefont {Emge}}, \bibinfo {author}
  {\bibfnamefont {P.}~\bibnamefont {H\"{o}hn}}, \bibinfo {author}
  {\bibfnamefont {T.}~\bibnamefont {Hughbanks}}, \ and\ \bibinfo {author}
  {\bibfnamefont {Y.}~\bibnamefont {Tian}},\ }\bibfield  {title} {\enquote
  {\bibinfo {title} {Crystal structure, electrical transport, and magnetic
  properties of niobium monophosphide},}\ }\href {\doibase 10.1021/ic950826f}
  {\bibfield  {journal} {\bibinfo  {journal} {Inorg. Chem.}\ }\textbf {\bibinfo
  {volume} {35}},\ \bibinfo {pages} {845--849} (\bibinfo {year}
  {1996})}\BibitemShut {NoStop}%
\bibitem [{\citenamefont {Saparov}\ \emph {et~al.}(2012)\citenamefont
  {Saparov}, \citenamefont {Mitchell},\ and\ \citenamefont
  {Sefat}}]{Saparov_SST_2012}%
  \BibitemOpen
  \bibfield  {author} {\bibinfo {author} {\bibfnamefont {B.}~\bibnamefont
  {Saparov}}, \bibinfo {author} {\bibfnamefont {J.~E.}\ \bibnamefont
  {Mitchell}}, \ and\ \bibinfo {author} {\bibfnamefont {A.~S.}\ \bibnamefont
  {Sefat}},\ }\bibfield  {title} {\enquote {\bibinfo {title} {Properties of
  binary transition-metal arsenides (\textit{{T}}{A}s)},}\ }\href@noop {}
  {\bibfield  {journal} {\bibinfo  {journal} {Supercond. Sci. Technol.}\
  }\textbf {\bibinfo {volume} {25}},\ \bibinfo {pages} {084016} (\bibinfo
  {year} {2012})}\BibitemShut {NoStop}%
\bibitem [{Cry(2014)}]{CrysAlisPro}%
  \BibitemOpen
  \href@noop {} {}\bibinfo {howpublished} {Agilent Technologies UK Ltd.,
  Oxford, UK} (\bibinfo {year} {2014}),\ \bibinfo {note} {{A}gilent
  {T}echnologies {C}rys{A}lis{P}ro, version 1.171.37.33}\BibitemShut {NoStop}%
\bibitem [{\citenamefont {Betteridge}\ \emph {et~al.}(2003)\citenamefont
  {Betteridge}, \citenamefont {Carruthers}, \citenamefont {Cooper},
  \citenamefont {Prout},\ and\ \citenamefont {Watkin}}]{Crystals}%
  \BibitemOpen
  \bibfield  {author} {\bibinfo {author} {\bibfnamefont {P.~W.}\ \bibnamefont
  {Betteridge}}, \bibinfo {author} {\bibfnamefont {J.~R.}\ \bibnamefont
  {Carruthers}}, \bibinfo {author} {\bibfnamefont {R.~I.}\ \bibnamefont
  {Cooper}}, \bibinfo {author} {\bibfnamefont {K.}~\bibnamefont {Prout}}, \
  and\ \bibinfo {author} {\bibfnamefont {D.~J.}\ \bibnamefont {Watkin}},\
  }\bibfield  {title} {\enquote {\bibinfo {title} {\textit{{CRYSTALS}} version
  12: software for guided crystal structure analysis},}\ }\href {\doibase
  10.1107/S0021889803021800} {\bibfield  {journal} {\bibinfo  {journal} {J.
  Appl. Cryst.}\ }\textbf {\bibinfo {volume} {36}},\ \bibinfo {pages} {1487}
  (\bibinfo {year} {2003})},\ \bibinfo {note} {we used version
  14.5481}\BibitemShut {NoStop}%
\bibitem [{\citenamefont {Villars}\ and\ \citenamefont {Cenzual}()}]{Pearson}%
  \BibitemOpen
  \bibfield  {author} {\bibinfo {author} {\bibfnamefont {P.}~\bibnamefont
  {Villars}}\ and\ \bibinfo {author} {\bibfnamefont {K.}~\bibnamefont
  {Cenzual}},\ }\href@noop {} {\enquote {\bibinfo {title} {{P}earson's
  {C}rystal {D}ata: {C}rystal {S}tructure {D}atabase for {I}norganic
  {C}ompounds, {R}elease 2014/15},}\ }\bibinfo {howpublished} {ASM
  International, Materials Park, Ohio, USA}\BibitemShut {NoStop}%
\bibitem [{\citenamefont {Bergerhoff}\ and\ \citenamefont
  {Brown}(1987)}]{ICSD}%
  \BibitemOpen
  \bibfield  {author} {\bibinfo {author} {\bibfnamefont {G.}~\bibnamefont
  {Bergerhoff}}\ and\ \bibinfo {author} {\bibfnamefont {I.~D.}\ \bibnamefont
  {Brown}},\ }\enquote {\bibinfo {title} {Inorganic crystal structure
  database},}\ in\ \href@noop {} {\emph {\bibinfo {booktitle} {Crystallographic
  Databases}}},\ \bibinfo {editor} {edited by\ \bibinfo {editor} {\bibfnamefont
  {F.~H.}\ \bibnamefont {Allen}}, \bibinfo {editor} {\bibfnamefont
  {G.}~\bibnamefont {Bergerhoff}}, \ and\ \bibinfo {editor} {\bibfnamefont
  {R.}~\bibnamefont {Sievers}}}\ (\bibinfo  {publisher} {International Union of
  Crystallography},\ \bibinfo {address} {Chester, UK},\ \bibinfo {year}
  {1987})\ pp.\ \bibinfo {pages} {77--95}\BibitemShut {NoStop}%
\end{thebibliography}%

\end{document}